\documentclass[pre,twocolumn]{revtex4-2}
\usepackage{graphicx}
\usepackage{amsopn}
\usepackage{amsfonts}
\usepackage{amsmath}
\usepackage{latexsym}
\usepackage{bm}
\usepackage{amssymb}
\usepackage{color}

\begin{document}

\title{Perturbation theory for phase correlations of a light wave \\
propagating in a turbulent medium}

\author{I.V. Kolokolov$^{1,2}$ and V.V. Lebedev$^{1,2}$}

\affiliation{$^1$ Landau Institute for Theoretical Physics, RAS, \\
142432, Chernogolovka, Semenova 1A, Moscow region, Russia; \\
$^2$ NRU Higher School of Economics, \\
101000, Myasnitskaya 20, Moscow, Russia. }

\date{\today}

\begin{abstract}

We theoretically investigate the correlation functions of the phase of a light wave propagating through a turbulent medium. We use an equation for the logarithm of a wave packet envelope, which includes a second-order nonlinear term. Based on this equation, we develop a diagrammatic technique to calculate corrections to the correlation function obtained in the linear approximation. We calculate the first corrections determined by one-loop diagrams and find its asymptotic behaviors. Some non-perturbative conclusions are made using the symmetry properties of the equation. These results allow us to conclude that the applicability condition for the perturbation theory is the smallness of the Rytov dispersion, $\sigma_R^2$, and this condition holds uniformly over the distances between observation points.

\end{abstract}


\maketitle

\section{Introduction}
\label{sec:intoduction}

This paper focuses on the theoretical analysis of light beam propagation through a turbulent medium, such as the Earth's atmosphere. Turbulence causes fluctuations in the refractive index, leading to distortions in a light beam. Observations have revealed that the strength of these distortions depends on the path taken by the beam and its direction relative to the Earth's surface. For relatively short distances, the distortions are minor, and this situation is known as the weak scintillation regime. However, over longer distances, the light beam passes into the strong scintillation regime, splitting into separate speckles.

The main ideas used to explain the peculiarities of light propagation in a turbulent medium can be traced back to the classic works of Kolmogorov and Obukhov, see Refs. \cite{Kolm411,Kolm412,Obukhov41}. In these studies, based on the concept of energy cascade, scaling laws have been established that characterize fluctuations of velocity within the inertial range of scales of turbulence. Subsequently, Obukhov \cite{Obuk49} and Corrsin \cite{Corr51} expanded this approach to include fluctuations of a passive scalar, such as temperature or impurity density, in a turbulent medium. These findings can also be used to describe the statistical properties of fluctuations of the refractive index, as it was noted in Ref. \cite{Obuk53}.

The problems related to the light propagation in a turbulent medium were intensively studied in the second half of the twentieth century. The results of the investigations are outlined in Refs. \cite{Tat67,Goodman,Tat75,Tat751,StrohbehnBook1978,AP98}. They concern mainly typical events and their statistical properties. Besides, some results for the strong scintillation regime were also obtained, see, e.g., Ref. \cite{Char94}. Recently, interest in the problem has been renewed mainly due to the increasing possibilities of numerical simulations, which allow one to obtain detailed information about light propagation in a turbulent medium, see Refs. \cite{Vorontsov2010,Vorontsov2011,Lachinova2016,Lushnikov2018,Aks19}. The systematic investigation of the tails of the probability density function of the light intensity is reported in our works \cite{KL23,KL24,KL25,KL25UFN}.

Distortions caused by atmospheric turbulence can be partially corrected using a technique called adaptive optics, see, e.g., Refs. \cite{Beck93,Tys10}. In Ref. \cite{BKLS24}, we have proposed a new method for measuring the Fried parameter, $r_0$, see Ref. \cite{Fried75}. The parameter is a crucial characteristic of radiation in the turbulent atmosphere determining the requirements for spatial resolution in adaptive optics systems. The idea of Ref. \cite{BKLS24} is in exploiting the off-diagonal component of the phase gradient correlation function, which can be measured, say, using a Shack-Hartmann wavefront sensor \cite{Shack01}.

The expressions for the correlation functions of the phase gradients presented in Ref. \cite{BKLS24} were derived for the case where the separation $r$ between the observation points is much larger than the radius of the first Fresnel zone. The results reported in Ref. \cite{KLS25} expand the approach to arbitrary separations $r$. The explicit expressions presented in Refs. \cite{BKLS24} and \cite{KLS25} were derived from the linearized equation for the logarithm of the envelope of the wave packet. A necessary condition for the application of this approach is the smallness of the Rytov variance $\sigma_R^2$ \cite{Tat67}. Direct numerical simulations, as described in Ref. \cite{Nemts}, have shown that the approximation performs well in the weak scintillation regime.

However, Refs. \cite{Varva,Baraban,Tat751} state that the inequality $\sigma_R^2 \ll 1$ may not be sufficient to accurately calculate the correlation function of the phase for all distances $r$ using perturbation theory. The reason for this is that there may be terms in the perturbation series that are singular at $r\to 0$ or $r\to \infty$, which violates the uniformity of the approximation. Indeed, the expansion in $\sigma_R^2$ of the correlation functions of the envelope $\Psi$ ceases to be valid for sufficiently large $r$.

To establish the applicability conditions of the perturbation theory for the phase correlations, we have developed a perturbation theory that allows us to calculate corrections to the expressions due to non-linearity of the equation for the logarithm of the envelope. The corrections can be presented by Feynman diagrams of various orders. Here, we discuss mainly the results of our calculations for the first set of corrections, which were determined by one-loop Feynman diagrams. Some evaluations concerning higher order corrections are also analyzed. The main issue with the calculations is that the perturbation series for the correlation functions of the logarithm of the envelope is uniform in $\sigma_R^2$ unlike the perturbation series for the envelope itself. It is worth noting that, for this statement to hold true for the phase susceptibility function, partial summation of the perturbation expansion should be done in certain limiting cases.

The structure of our paper is as follows. Section \ref{sec:mainrel} contains the main relations concerning the light propagation in a turbulent medium, including fluctuations of the refractive index, the envelope of the wave packet and its logarithm. In Section \ref{sec:perturb} we explain the construction of the perturbation series for the correlation function of the logarithm of the envelope. Section \ref{sec:first} is devoted to analyzing the first corrections to the pair correlation functions and to the Green's function. In Section \ref{sec:largesc} we, based on symmetry properties of the problem, analyze non-perturbatively the role of large-scale fluctuations of the refractive index. In Section \ref{sec:conclusion}, we summarize our findings, draw main conclusions, and propose possible extensions to our research. Technical details of our calculations are presented in Appendices.

\section{Main relations}
\label{sec:mainrel}

We consider a monochromatic light wave traveling through a turbulent medium, such as the atmosphere. The wave packets of these waves can be described by their envelope, which determines the overall shape of the wave on scales larger than the wavelength. The envelope $\Psi$ is a complex function of time $t$, of two-dimensional radius-vector $\bm r=(x,y)$ perpendicular to the direction of the wave packet propagation, and of the coordinate $z$ along this direction.

The envelope $\Psi$ is controlled by the following parabolic equation
\begin{equation}
i \partial_z \Psi +\frac{1}{2k_0}\nabla^2 \Psi
+k_0 \nu  \Psi=0.
\label{gain1}
\end{equation}
Here $k_0$ is the wave vector of the carrying wave, and $\nu$ designates a fluctuation of the refraction index. The vector operator $\nabla$ in Eq. (\ref{gain1}) designates the coordinate gradient in the direction perpendicular to the $Z$-axis, $\nabla=(\partial_x,\partial_y)$. The equation (\ref{gain1}) implies that the envelope, $\Psi$, adapts simultaneously to the state of the medium, which is described by the field $\nu$. This property is justified by the speed of light being very high.

The equation (\ref{gain1}) is linear, so we have neglected any non-linear optical phenomena, such as those associated with the Kerr effect. This approximation assumes that the amplitude of the light wave is small enough. Strong light waves, which can exhibit non-linear behavior, are a topic that is beyond the scope of this paper.

Statistical properties of the refraction index $\nu$ can be characterized by its second order structure function. In the inertial range of turbulence the structure function follows a power law
\begin{eqnarray}
\langle [\nu  (\bm r_1,z_1) -\nu  (\bm r_2,z_2)]^2\rangle
=C_n^2 (r^2+z^2)^{\mu/2},
\label{KolmOb}
\end{eqnarray}
characteristic of a passive scalar. Here and below the angular brackets denote time averaging. In the expression (\ref{KolmOb}), the quantity $C_n^2$ determines the strength of the fluctuations of the refractive index, $\bm r=\bm r_1-\bm r_2$, $z=z_1-z_2$ and $\mu$ is a scaling index. The expression (\ref{KolmOb}) suggests that the turbulent fluctuations in the inertial range are statistically homogeneous and isotropic, which is consistent with experimental observations \cite{Monin,Frisch}.

For the Obukhov-Corrsin spectrum \cite{Obuk49,Corr51} $\mu=2/3$. This exponent is commonly observed in turbulent media including atmosphere \cite{Monin,Frisch}. Below, for the sake of generality, we treat $\mu$ to be a parameter that falls between $0$ and $1$, $0<\mu<1$.

We assume that the propagation path of the wave packet is much larger than all its characteristic lateral scales. Then, the field $\nu$ can be treated as a quantity which is short correlated along the $Z$-axis. The property motivates the approximation
\begin{eqnarray}
\langle \nu(\bm r_1,z_1) \nu(\bm r_2,\bm z_2)\rangle=
\delta(z_1-z_2) C_n^2 {\mathcal A}(\bm r),
\label{delta}
\end{eqnarray}
where $\bm r=\bm r_1-\bm r_2$. The function $\mathcal A$ in Eq. (\ref{delta}) is determined by the spectrum of the fluctuations of the refractive index. The factor $C_n^2$ in Eq. (\ref{delta}) can be thought as a gradually varying function of $z$.

Since $\nu$ appears in observable quantities through integrals over $z$, it can be assumed to have Gaussian statistics, due to the central limit theorem, see, e.g., Ref. \cite{Feller}. In this case, the pair correlation function (\ref{delta}) fully determines the statistical properties of the field $\nu$. This function is involved in all calculations related to the statistical properties of the light wave distortions.

It is instructive to deal with the Fourier transform $\tilde{\mathrm A}$ of the function ${\mathcal A}$:
\begin{eqnarray}
{\mathcal A}(\bm r) =
\int \frac{d^2 q}{(2\pi)^2}
\exp(i \bm q \bm r)
\tilde{\mathcal A}(\bm q) .
\label{afo1}
\end{eqnarray}
For the inertial region of scales, one finds
\begin{eqnarray}
\tilde{\mathcal A}(\bm q)=2\pi
\Gamma(2+\mu)\sin(\pi \mu/2)
q^{-3-\mu}.
\label{afo2}
\end{eqnarray}
The factor at $q^{-3-\mu}$ is dictated by the expression (\ref{KolmOb}) for the structure function of $\nu$. The integral (\ref{afo1}) diverges at small $q$. The divergence should be cut at the inverse outer scale of turbulence. For example, one can use von Karman spectrum \cite{Karman48} instead of Eq. (\ref{afo2}).

Calculating the integral (\ref{afo1}) one obtains for scales from the inertial interval
\begin{eqnarray}
{\mathcal A}(r)={\mathcal A}_0
- \frac{\sqrt \pi}{2} \frac{\Gamma(-\mu/2-1/2)}{\Gamma(-\mu/2)}  r^{1+\mu}
+{\mathcal M} r^2.
\label{corrfa2}
\end{eqnarray}
The contributions with the constants ${\mathcal A}_0$ and $\mathcal M$ in Eq. (\ref{corrfa2}) are caused by the turbulent fluctuations at the outer scale of turbulence. Designating the scale as $L_0$ we find the following estimates: ${\mathcal A}_0\sim L_0^{1+\mu}$ and ${\mathcal M} \sim L_0^{\mu-1}$. If the term with the factor ${\mathcal M}$ is neglected then the expression (\ref{corrfa2}) leads to Eq. (\ref{KolmOb}).

The expression for the second order correlation function of the derivatives of the refractive index follows from Eq. (\ref{corrfa2}):
\begin{equation}
-\partial_\alpha \partial_\beta {\mathcal A}(\bm r)
=S_{\alpha\beta}(\bm r)-2{\mathcal M}\delta_{\alpha\beta}.
\label{parpara}
\end{equation}
Note that the constant ${\mathcal A}_0$ dropped from the expression (\ref{parpara}). The function $S_{\alpha\beta}$ in Eq. (\ref{parpara}) is written as
\begin{eqnarray}
S_{\alpha\beta}(\bm r)
=-\sqrt\pi \frac{\Gamma(1/2-\mu/2)}{\Gamma(-\mu/2)} r^{\mu-1}
\nonumber \\
\left[\delta_{\alpha\beta}
-(1-\mu) \frac{r_\alpha r_\beta}{r^{2}}\right]
\label{parpamu}
\end{eqnarray}
The quantity tends to infinity as $r\to0$. The divergence is cut off at the internal scale of turbulence.

The degree of the light signal distortion is usually characterized by the Rytov variance $\sigma_R^2$. For the Obukhov-Corrsin spectrum where $\mu=2/3$, and a distorted plane wave, the definition of the Rytov variance is
\begin{equation}
\sigma_R^2=2.25 \, k_0^{7/6} z^{5/6}
\int_0^z d\zeta\, (1-\zeta/z)^{5/6} C_n^2(\zeta),
\label{pla2}
\end{equation}
see, e.g., Ref. \cite{Spencer2021}. For a homogeneous medium one finds
\begin{equation}
\sigma_R^2=1.23\,  C_n^2 k_0^{7/6} z^{11/6},
\label{rytovd}
\end{equation}
after taking the integral in Eq. (\ref{pla2}).

Considering $\mu$ as an arbitrary parameter, we generalize the definition (\ref{rytovd}) as
\begin{equation}
\sigma_R^2
\sim C_n^2 k_0^{3/2-\mu/2}z^{3/2+\mu/2}.
\label{res67}
\end{equation}
We do not specify the factor in the definition (\ref{res67}) implying that for $\mu=2/3$ it coincides with one in Eq. (\ref{rytovd}).

\subsection{Logarithm of the envelope}

Let us pass from the envelope $\Psi$ to its logarithm $\ln \Psi$. The phase $\varphi$ of the envelope $\Psi$ is the imaginary part of $\ln \Psi$, whereas the real part of $\ln \Psi$ determines the absolute value $|\Psi|$ of the envelope. The equation (\ref{gain1}) leads to the following equation for the logarithm of $\Psi$
\begin{equation}
\partial_z \ln\Psi
-\frac{i}{2k_0}\nabla^2 \ln\Psi
-\frac{i}{2k_0}(\nabla \ln\Psi)^2
-i k_0 \nu =0.
\label{logpsi}
\end{equation}
Note that, unlike the linear equation (\ref{gain1}), the equation (\ref{logpsi}) is nonlinear, it has a second-order nonlinearity. Besides, the random field $\nu$ enters the equation (\ref{logpsi}) as an addition whereas in Eq. (\ref{gain1}) it was a factor at $\Psi$.

The logarithm of the envelope, $\ln \Psi$, includes both the contribution from the free propagation of the envelope (without random diffraction) and the fluctuating contribution caused by $\nu$. It is convenient to exclude the first contribution passing to the quantity $\eta=\ln (\Psi/\Psi_0)$, where $\Psi_0$ is the envelope for the free propagation of the wave. The envelope $\Psi_0$ is a solution of the equation (\ref{logpsi}) for $\nu =0$, being dependent on the initial envelope profile. The dimensionless quantity $\eta$ characterizes just the distortions of the profile $\Psi_0$.

The equation for the fluctuating quantity $\eta$ can be derived from the equation (\ref{logpsi}), it is written as
\begin{eqnarray}
\partial_z \eta
-\frac{i}{2k_0}\nabla^2 \eta
-\frac{i}{k_0} \nabla \ln\Psi_0 \nabla\eta
\nonumber \\
=i k_0 \nu +\frac{i}{2k_0}(\nabla \eta)^2.
\label{log1}
\end{eqnarray}
Note that the equation is, generally, spatially inhomogeneous due to the presence of the term containing $\Psi_0$. The equation (\ref{log1}) needs to be solved with a zero initial condition, $\eta=0$ at $z=0$. Taking the equation complex conjugated to Eq. (\ref{log1}) one finds the equation for $\eta^\star$ where star designates complex conjugation.

The main objects interesting for us are the pair correlation functions
\begin{eqnarray}
\varPhi(\bm r_1,z_1;\bm r_2,z_2)
=\langle \eta(\bm r_1,z_1)
\eta(\bm r_2,z_2) \rangle,
\label{pairphi} \\
\varXi(\bm r_1,z_1;\bm r_2,z_2)
=\langle \eta(\bm r_1,z_1)
\eta^\star(\bm r_2,z_2) \rangle,
\label{pairxi}
\end{eqnarray}
related to the field $\eta$. Here, angular brackets denote time averaging, or, in a theoretical context, averaging over the statistics of $\nu$. The pair correlation function of the phase fluctuations $\varphi$ of the envelope $\Psi$ is expressed as
\begin{equation}
\langle \varphi(\bm r_1,z_1) \varphi(\bm r_2,z_2) \rangle
=-\frac{1}{4}\left(\varPhi+\varPhi^\star-\varXi -\varXi^\star\right),
\label{log3}
\end{equation}
where all the functions in the right-hand side of the equation have the same arguments $\bm r_1, z_1; \bm r_2, z_2$.

The Shack-Hartmann method \cite{Shack01} deals with the pair correlation function of the phase derivatives, which is the following matrix $2\times2$
\begin{equation}
Q_{\alpha\beta}=
\left\langle \frac{\partial}{\partial r_{1\alpha}}\varphi(\bm r_1,z_1)
\frac{\partial}{\partial r_{2\beta}}
\varphi(\bm r_2,z_2) \right\rangle.
\label{qalbe}
\end{equation}
In accordance with Eq. (\ref{log3}), the correlation function (\ref{qalbe}) is expressed via the derivatives of the functions $\varXi,\varPhi$:
\begin{eqnarray}
Q_{\alpha\beta}=\frac{1}{2}\mathrm{Re}\,
(\varXi_{\alpha\beta}-\varPhi_{\alpha\beta}),
\label{qalbe2} \\
\varXi_{\alpha\beta}=\frac{\partial^2\varXi}{\partial r_{1\alpha}\partial r_{2\beta}}, \quad
\varPhi_{\alpha\beta}=\frac{\partial^2\varPhi}{\partial r_{1\alpha}\partial r_{2\beta}}.
\label{qalbe3}
\end{eqnarray}
The objects $\varXi_{\alpha\beta}$, $\varPhi_{\alpha\beta}$ (\ref{qalbe3}) are symmetric matrices $2\times2$.

Besides the correlation functions (\ref{pairphi},\ref{pairxi}) one is interested in the susceptibility of the system. It determines the average value of the field $\eta$, $\langle \eta \rangle$, caused by an addition of the infinitesimal deterministic contribution $\delta \nu$ to the fluctuating field $\nu$:
\begin{eqnarray}
\langle \eta(\bm r,z) \rangle
=i k_0\int d^2r_1 d\zeta\,
G(\bm r,z;\bm r_1,\zeta)
\delta \nu (\bm r_1,\zeta).
\label{agra1}
\end{eqnarray}
The susceptibility $G$ figuring in Eq. (\ref{agra1}) can be called as the Green's function.

\subsection{Linear approximation for $\eta$}
\label{subsec:zeroth}

We further develop a perturbation theory based on the nonlinear term in Eq. (\ref{log1}). The zeroth approximation of the perturbation series corresponds to neglecting the nonlinear term in Eq. (\ref{log1}). For the cases of the perturbed plane wave and the perturbed spherical wave the zeroth correlation functions  $\varXi_0,\varPhi_0$ are found in Ref. \cite{KLS25}. Here, we briefly recall the derivation of the zeroth-order approximation with an eye toward its generalization for higher orders in the perturbation series.

 Neglecting the nonlinear term in Eq. (\ref{log1}) we find the linear equations which can be solved in terms of $\nu$ to obtain
\begin{equation}
\eta(\bm r,z)=i k_0
\int d^2r_1 d\zeta\, G_0(\bm r,z;\bm r_1,\zeta)
\nu(\bm x,\zeta),
\label{responce}
\end{equation}
where the function $G_0$ satisfies the equation
\begin{eqnarray}
\left(\partial_z -\frac{i}{2k_0}\nabla^2\right)
G_0(\bm r,z;\bm r_1,\zeta)
\nonumber \\
-\frac{i}{k_0} \nabla \ln\Psi_0(\bm r,z) \nabla
G_0(\bm r,z;\bm r_1,\zeta)
\nonumber \\
= \delta(\bm r-\bm r_1)\delta(z-\zeta),
\label{greenlog}
\end{eqnarray}
and is equal to zero if $\zeta>z$ due to causality. If $z\to \zeta+0$ then $G_0\to \delta(\bm r-\bm r_1)$. Comparing the relation (\ref{responce}) to Eq. (\ref{agra1}) we conclude that $G_0$ gives the zeroth approximation for the Green's function $G$.

The correlation functions $\varPhi$, $\varXi$ (\ref{pairphi},\ref{pairxi}) in the zeroth approximation can be extracted from the relations (\ref{delta},\ref{responce}). One finds
\begin{eqnarray}
\varPhi_0(\bm r_1,z_1; \bm r_2,z_2)
=-k_0^2\int d^2 r_3\, d^2 r_4\, d\zeta\, C_n^2(\zeta)
\nonumber \\
G_0(\bm r_1,z_1;\bm r_3,\zeta)
G_0(\bm r_2,z_2;\bm r_4,\zeta)
{\mathcal A}(\bm r_3-\bm r_4),
\label{respo1}
\end{eqnarray}
and
\begin{eqnarray}
\varXi_0(\bm r_1,z_1; \bm r_2,z_2)
=k_0^2\int d^2 r_3\, d^2 r_4\, d\zeta\, C_n^2(\zeta)
\nonumber \\
G_0(\bm r_1,z_1;\bm r_3,\zeta)
G_0^\star(\bm r_2,z_2;\bm r_4,\zeta)
{\mathcal A}(\bm r_3-\bm r_4),
\label{respo2}
\end{eqnarray}
where the integration over $\zeta$ is performed from $0$ to minimum of $z_1,z_2$.

Further we restrict ourselves to the case of the initial plane wave perturbed by the refractive index fluctuations. Then the system is homogeneous in average in the lateral direction and the correlation functions depend on the differences of the lateral coordinates. Details of the derivations can be Found in Appendix \ref{sec:zeroth}.

For the initial plane wave $\Psi_0$ is independent of coordinates. Then the term with $\Psi_0$ drops from Eq. (\ref{greenlog}) and one immediately finds its solution, which is
\begin{eqnarray}
G_0(\bm r,z;\bm r_1,\zeta)=\frac{k_0 \theta(z-\zeta)}{2\pi i (z-\zeta)}
\exp\left[ i\frac{k_0 (\bm r-\bm r_1)^2}{2(z-\zeta)}\right],
\label{logpl}
\end{eqnarray}
where $\theta$ is the Heaviside step function. Note, that the expression (\ref{logpl}) is homogeneous in space. Its Fourier transform in terms of the difference $\bm r-\bm r_1$ is equal to
\begin{equation}
\tilde G_0(\bm k,z,\zeta)=\theta(z-\zeta) \exp\left[ -\frac{i(z-\zeta)}{2k_0}k^2 \right],
\label{log4}
\end{equation}
where $\bm k$ is the $2d$ wave vector.

If $z_1=z_2$ then the function $\varXi_0$ (\ref{respo2}) is reduced to
\begin{eqnarray}
\varXi_{0}(\bm r,z;\bm 0,z)
=k_0^2 \int_0^z d\zeta\, C_n^2 {\mathcal A}(\bm r).
\label{vix1}
\end{eqnarray}
It can be easily checked by performing Fourier transform in $\bm r_1-\bm r_2$ of the expression (\ref{respo2}) and then by using Eq. (\ref{log4}). The quantity $\varXi_{0\alpha\beta}$ is written as
\begin{eqnarray}
\varXi_{0\alpha\beta}(\bm r,z;\bm 0,z)
=k_0^2 \int_0^z d\zeta\, C_n^2 S_{\alpha\beta}(\bm r).
\label{vix2}
\end{eqnarray}
as a consequence of Eq. (\ref{vix1}). We neglected in Eq. (\ref{vix2}) the term involving $\mathcal{M}$ regarding the external length of turbulence $L_0$ to be large enough. Therefore $\varXi_{0\alpha\beta}(\bm r,z;\bm 0,z)$ can be estimated as
\begin{eqnarray}
\varXi_{0\alpha\beta}(\bm r,z;\bm 0,z)\sim
k_0^2 C_n^2 z r^{\mu-1},
\label{vix3}
\end{eqnarray}
it diverges as $r\to 0$.

Unlike $\varXi_{0\alpha\beta}(\bm r,z;\bm 0,z)$, the quantity $\varPhi_{0\alpha\beta}(\bm r,z;\bm 0,z)$ remains finite as $r\to0$ \cite{KLS25}. We find the estimate
\begin{eqnarray}
\varPhi_{0\alpha\beta}(\bm r,z;\bm 0,z)\sim
C_n^2 k_0^{5/2-\mu/2}z^{1/2+\mu/2},
\label{vix4}
\end{eqnarray}
correct for $k_0 r^2\ll z$, see details in Appendix \ref{sec:zeroth}. In the opposite case $k_0 r^2\gg z$ we obtain
\begin{eqnarray}
\varXi_0(\bm r_1,z;\bm r_2,z)
=-\varPhi_0(\bm r_1,z;\bm r_2,z),
\label{larger} \\
\varXi_{0\alpha\beta}(\bm r_1,z;\bm r_2,z)
=-\varPhi_{0\alpha\beta}(\bm r_1,z;\bm r_2,z) \propto r^{\mu-1},
\label{lig3}
\end{eqnarray}
see Eqs. (\ref{vix1},\ref{vix2}). We exploited the approximation (\ref{lig3}) in Ref. \cite{BKLS24}.

\section{Perturbation theory}
\label{sec:perturb}

We begin developing a perturbation theory for calculating the correlation functions $\varPhi,\varXi,G$ defined by Eqs. (\ref{pairphi},\ref{pairxi},\ref{agra1}). The perturbation theory can be constructed in accordance with the general scheme proposed in Ref. \cite{MSR73}. A comprehensive description of the technique can be found in Ref \cite{Penco}. Note that this type of technique was first proposed by Wyld \cite{Wyld61} in the context of turbulence.

It is convenient for us to represent the correlation function of $\eta$ in terms of functional integrals, see, e.g., Ref. \cite{Popov76}. The weight in the functional integrals is $\exp(iS)$, like in quantum field theory, see, e.g., Ref. \cite{Weinberg}. Here, $S$ is the effective action, which is a functional of the fields relevant to the problem at hand. For our problem, the effective action $S$ can be found from Eqs. (\ref{delta},\ref{log1}), like it is described in Ref. \cite{HRS17}.

As a result, we find the effective action which is the sum $S=S_0+S_{int}$, where
\begin{eqnarray}
iS_0=\int d^2r\, dz\,\left\{
i \bar p \partial_z \eta +i p \partial_z \eta^\star
+ \frac{\bar p}{2k_0}\nabla^2 \eta
-\frac{p}{2k_0}\nabla^2 \eta^\star
\right.
\nonumber \\ \left.
+\frac{\bar p}{k_0}\nabla \ln \Psi_0 \nabla\eta
-\frac{p}{k_0}\nabla \ln \Psi_0^\star \nabla\eta^\star
+k_0 (\bar p-p) \delta\nu
\right\}
\nonumber \\
+\frac{k_0^2}{2}\int d^2 r_1 d^2 r_2 dz
(p_1-\bar p_1) C_n^2 {\mathcal A}(\bm r_1-\bm r_2)
(p_2-\bar p_2) , \qquad
\label{action}
\end{eqnarray}
and
\begin{eqnarray}
i S_{int}=\frac{1}{2k_0}\int d^2r\, dz\,\left[
\bar p (\nabla \eta)^2
-p (\nabla \eta^\star)^2 \right]. \qquad
\label{interaction}
\end{eqnarray}
Here $p$ and $\bar p$ are auxiliary fields. The fields $p_1,\bar p_1$ and $p_2,\bar p_2$ in Eq. (\ref{action}) are functions of $\bm r_1,z$ and $\bm r_2,z$, respectively. We included into the expression for the effective action the term with the infinitesimal deterministic addition $\delta\nu$ to the refraction index.

The correlation functions of the fields $\eta,\eta^\star,p,\bar p$ can be written as functional integrals over the fields. The pair correlation functions $\varPhi,\varXi$ are written as
\begin{eqnarray}
\varPhi=\langle \eta(\bm r,z)\eta(\bm r_1,\zeta)\rangle
=\int D\eta\, D\eta^\star\, Dp\, D\bar p
\nonumber \\
\exp(iS) \eta(\bm r,z)\eta(\bm r_1,\zeta),
\label{action0}
\end{eqnarray}
and
\begin{eqnarray}
\varXi=\langle \eta(\bm r,z)\eta^\star(\bm r_1,\zeta)\rangle
=\int D\eta\, D\eta^\star\, Dp\, D\bar p
\nonumber \\
\exp(iS) \eta(\bm r,z)\eta^\star(\bm r_1,\zeta).
\label{actionn}
\end{eqnarray}
Analogously higher order correlation functions of $\eta$, $\eta^\star$ can be presented.

One can relate the Green's function $G$ to the pair correlation function containing an auxiliary field. Writing
\begin{equation}
\langle \eta \rangle
=\int D\eta\, D\eta^\star\, Dp\, D\bar p
\exp(iS) \eta,
\nonumber
\end{equation}
and then expanding $\exp(iS)$ in $\delta\nu$, one finds in the linear approximation
\begin{eqnarray}
\langle \eta(\bm r,z) \rangle
=\int d^2 r_1\, d\zeta\,
\int D\eta\, D\eta^\star\, Dp\, D\bar p
\exp(iS)
\nonumber \\
k_0 \eta(\bm r,z)
\bar p (\bm r_1,\zeta) \delta \nu(\bm x,\zeta).
\nonumber
\end{eqnarray}
Comparing the relation to Eq. (\ref{agra1}) we find
\begin{eqnarray}
\langle \eta(\bm r,z) \bar p(\bm r_1,\zeta)\rangle
= i G(\bm r,z; \bm r_1,\zeta),
\label{action2}
\end{eqnarray}
where
\begin{eqnarray}
\langle \eta(\bm r,z) \bar p(\bm r_1,\zeta)\rangle
= \int D\eta\, D\eta^\star\, Dp\, D\bar p
\nonumber \\
\exp(iS) \eta(\bm r,z) \bar p(\bm r_1,\zeta),
\label{action3}
\end{eqnarray}
in accordance with the general rules. Analogously
\begin{eqnarray}
\langle \eta^\star(\bm r,z) p(\bm r_1,\zeta)\rangle
= i G^\star(\bm r,z; \bm r_1,\zeta).
\label{action4}
\end{eqnarray}
The pair correlation functions $\langle \eta p \rangle$ and $\langle \eta^\star \bar p \rangle$ are zero. The pair correlation functions of the fields $p,\bar p$ are zero as well.

The zeroth approximation discussed in Subsection \ref{subsec:zeroth} corresponds to neglecting in the functional integrals for $\varPhi,\varXi,G$ the third order term (\ref{interaction}). Then the integrals are Gaussian and can be easily calculated. The results coincide with the expressions obtained in Subsection \ref{subsec:zeroth} by direct averaging over the statistics of $\nu$.

\subsection{Feynman diagrams}

\begin{figure}
\begin{center}
\includegraphics[width=0.8 \textwidth]{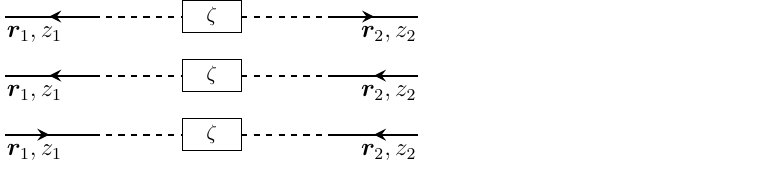}
\end{center}
\caption{Diagrammatic representation for $\varPhi_0,\varXi_0,\varPhi^\star_0$ in accordance with Eqs. (\ref{respo1},\ref{respo2}).}
 \label{fig:fdiagram1}
 \end{figure}

The perturbation theory for calculating $\varPhi,\varXi,G$ can be constructed as follows. One starts with the functional integrals, such as (\ref{action0},\ref{actionn},\ref{action3}), giving the expressions for $\varPhi,\varXi,G$. Expanding the factor
\begin{equation}
\exp(iS)=\exp(iS_0) \exp(iS_{int}),
\nonumber
\end{equation}
figuring in the integrals, into the series over $S_{int}$, one obtains terms which are determined by Gaussian integrals. They can be found explicitly in terms of $\varPhi_0,\varXi_0,G_0$ by using the Wick theorem \cite{Wick}.

Note that the interaction term $S_{int}$ (\ref{interaction}) contains gradients of $\eta$. Therefore the terms of the perturbation series are determined by the correlation functions of $\nabla\eta$, $\nabla\eta^\star$. That is why we analyze the correlation functions $\varPhi_{\alpha\beta}$, $\varXi_{\alpha\beta}$ besides the correlation functions $\varPhi$, $\varXi$.

The terms of the perturbation series correspond to Feynman diagrams. To read a diagram, one must specify its components. For the purpose we introduce graphs corresponding to the pair correlation functions $\varPhi_0,\varXi_0,\varPhi^\star_0$ determined by Eqs. (\ref{respo1},\ref{respo2}). The graphs are depicted in Fig. \ref{fig:fdiagram1}. Here the rectangles represent the factor $k_0^2 C_n^2 {\mathcal A}$, the combined dashed-solid line with the arrow pointing away from the dashed line represent the Green's function $G$, and the combined dashed-solid line, with the arrow pointing towards the dashed line, represents the Green's function $G^\star$.

\begin{figure}
\begin{center}
\includegraphics[width=0.75 \textwidth]{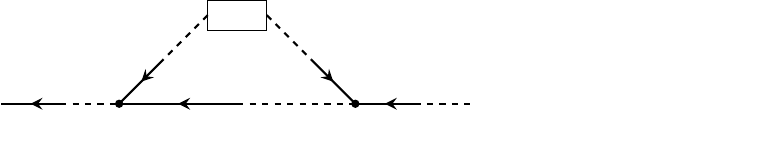}
\end{center}
\caption{Diagrammatic representation of a first-order (one-loop) correction to the Green's function $G$.}
 \label{fig:fdiagram2}
 \end{figure}

Now, we can consider corrections to $G_0, \varPhi_0, \varXi_0$ that are related to the account of the interaction term (\ref{interaction}) in the effective action. The first order correction to the Green's function $G$ is depicted in Fig. \ref{fig:fdiagram2}. It is obtained by expanding the factor $\exp(iS_{int})$ in Eq. (\ref{action3}) up to the second order and then applying the Wick theorem \cite{Wick}. The points in Fig. \ref{fig:fdiagram2} represent the triple vertices, which are determined by the interaction term (\ref{interaction}). We see that the correction is determined by an one-loop Feynman diagram.

\begin{figure}
\begin{center}
\includegraphics[width=0.65 \textwidth]{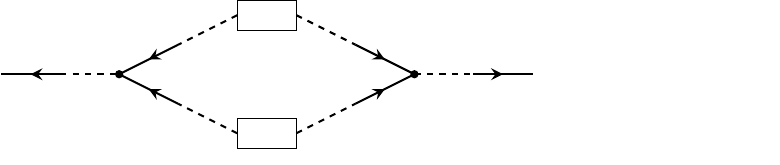}
\end{center}
\caption{Diagrammatic representation of a first-order (one-loop) correction to the pair correlation function $\varPhi$.}
 \label{fig:fdiagram3}
\end{figure}

\begin{figure}
\begin{center}
\includegraphics[width=0.65 \textwidth]{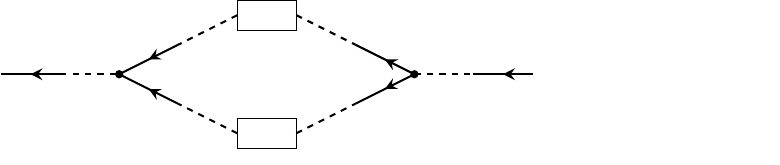}
\end{center}
\caption{Diagrammatic representation of a first-order (one-loop) correction to the pair correlation function $\varXi$.}
 \label{fig:fdiagram4}
 \end{figure}

Analogously, one can find first order corrections to the correlation functions $\varPhi,\varXi$. The corrections have some contributions. One of the first order contribution to $\varPhi$ is depicted in Fig. \ref{fig:fdiagram3}. An analogous structure has the diagram depicted in Fig. \ref{fig:fdiagram4} which represents a first-order contribution to $\varXi$. Both diagrams have one loop, similar to the case of the correction to the Green's function $G$, see Fig. \ref{fig:fdiagram2}.

The diagrams depicted in Figs. \ref{fig:fdiagram2}-\ref{fig:fdiagram4} can be treated as the first terms of the Dyson series. Based on the analogy with quantum electrodynamics, see, e.g., Ref. \cite{BLP82}, we can refer to the central (loop) part of Fig. \ref{fig:fdiagram2} as the first contribution to the self-energy function and the central (loop) parts of Figs. \ref{fig:fdiagram3},\ref{fig:fdiagram4} can be referred as the first contributions to the polarization functions.

\begin{figure}
\begin{center}
\includegraphics[width=0.65 \textwidth]{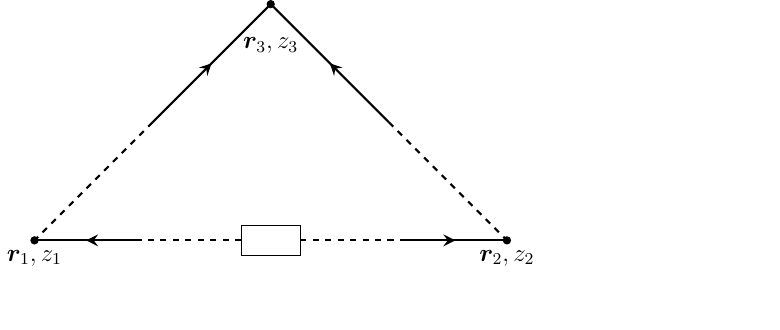}
\end{center}
\caption{Diagrammatic representation of a first-order (one-loop) correction to the triple vertex function.}
 \label{fig:fdiagram5}
 \end{figure}

Besides the corrections to the pair correlation functions one can consider corrections to the triple interaction vertex determined by the term (\ref{interaction}) in the effective action. A first order (one-loop) correction to the vertex is depicted in Fig. \ref{fig:fdiagram5}. Analogous to the diagrams shown in Figs. \ref{fig:fdiagram2}-\ref{fig:fdiagram3}, this term contains an additional factor $C_n^2$, compared to the zero-order expression.

One can think about higher order corrections to the correlation functions $G,\varPhi,\varXi$, which originates from higher order terms of the expansion of the factor $\exp(iS_{int})$ in the functional integrals giving the correlation functions. One can say that the higher the order of the perturbation series is, the more loops the corresponding diagrams have. Alternatively, one could say that the perturbation series is with respect to the parameter $C_n^2$.

\section{First-order corrections: analytics}
\label{sec:first}

Here we examine analytical expressions corresponding to the Feynman diagrams depicted in Figs. \ref{fig:fdiagram2}-\ref{fig:fdiagram4}. They determine the first order corrections $G_1$ to the zeroth order expression (\ref{logpl}), as well as the first order corrections $\varPhi_1,\varXi_1$ to the zeroth order expressions (\ref{respo1},\ref{respo2}) of the pair correlation functions.

The only first order correction to the Green's function is determined by the diagram depicted in Fig. \ref{fig:fdiagram2}. The corresponding analytical expression is
\begin{eqnarray}
G_1(\bm r_1,z_1;\bm r_2,z_2)
=-\frac{1}{k_0^2}\int d^2r_3 d^2r_4 d\zeta_1 d\zeta_2
\nonumber \\
G_0(\bm r_1,z_1;\bm r_3,\zeta_1)
\varPhi_{0\alpha\beta}(\bm r_3,\zeta_1;\bm r_4,\zeta_2)
\nonumber \\
\frac{\partial}{\partial r_{3\alpha}}G_0(\bm r_3,\zeta_1;\bm r_4,\zeta_2)
\frac{\partial}{\partial r_{4\beta}}G_0(\bm r_4,\zeta_2;\bm r_2,z_2),
\label{delt1}
\end{eqnarray}
where $\varPhi_{0\alpha\beta}$ is determined by Eq. (\ref{respo1}).

In the case of the initial plane wave the Green's function depend only on the difference of the coordinates $\bm r=\bm r_1-\bm r_2$. It is convenient to examine Fourier transform $\tilde G_1$ as a function of its wave vector $\bm k$. The corresponding analysis is done in Appendix \ref{sec:greenf1}. In the case $k^2\ll k_0/z$, we find the expression (\ref{dolt22}), therefore
\begin{equation}
\tilde G_1 \sim C_n^2 k^2 k_0^{1/2-\mu/2}z^{5/2+\mu/2}.
\nonumber
\end{equation}
In the opposite case $k^2\gg k_0/z$ we find the expression (\ref{gone1}), therefore
\begin{equation}
\tilde G_1(\bm k,z,z)\sim
C_n^2 z^{2+\mu}k^{1+\mu} k_0^{1-\mu}.
\nonumber
\end{equation}
If $z_1\neq z_2$ then $\tilde G_1$ contains an oscillating factor.

Now we pass to the pair correlation function $\varPhi$ (\ref{pairphi}). The first order correction $\varPhi_1$ to the expression (\ref{respo1}) consists of three terms
\begin{eqnarray}
\varPhi_1=\varPhi_1^{(1,3)}
+\varPhi_1^{(2,2)}+\varPhi_1^{(3,1)}.
\label{phi1}
\end{eqnarray}
The term $\varPhi_1^{(2,2)}$ is determined by the diagram depicted in Fig. \ref{fig:fdiagram3}. Two other terms, $\varPhi_1^{(1,3)}$ and $\varPhi_1^{(3,1)}$ are related to the correction to the Green's function determined by the diagram depicted in Fig. \ref{fig:fdiagram2}. They can be thought as the correction to the Green's functions presented in the diagram for $\varPhi_0$ depicted in Fig. \ref{fig:fdiagram1}.

The explicit expression for $\varPhi_1^{(2,2)}$ is
\begin{eqnarray}
\varPhi_1^{(2,2)}(\bm r_1,z_1;\bm r_2,z_2)=
\int d^2 r_3 d^2 r_4 d\zeta_3 d\zeta_4
\nonumber \\
G(\bm r_1,z_1;\bm r_3,z_3)
G(\bm r_2,z_2;\bm r_4,z_4)
\Pi(\bm r_3,z_3;\bm r_4,z_4), \quad
\label{phi122} \\
\Pi(\bm r_3,z_3;\bm r_4,z_4)
=-\frac{1}{2k_0^2}
\varPhi_{0,\alpha\beta}^2(\bm r_3,z_3;\bm r_4,z_4). \quad
\label{pi122}
\end{eqnarray}
The coefficients here are determined by the expressions (\ref{interaction},\ref{action2}). The contributions
$\varPhi_1^{(1,3)}$, $\varPhi_1^{(3,1)}$ are written as
\begin{eqnarray}
\varPhi_1^{(1,3)}(\bm r_1,z_1; \bm r_2,z_2)
=-k_0^2\int d^2 r_3\, d^2 r_4\,
d\zeta\, C_n^2(\zeta)
\nonumber \\
G_0(\bm r_1,z_1;\bm r_3,\zeta)
G_1(\bm r_2,z_2;\bm r_4,\zeta)
{\mathcal A}(\bm r_3-\bm r_4),
\label{respo11}
\end{eqnarray}
and
\begin{eqnarray}
\varPhi_1^{(3,1)}(\bm r_1,z_1; \bm r_2,z_2)
=-k_0^2\int d^2 r_3\, d^2 r_4\, d\zeta\, C_n^2(\zeta)
\nonumber \\
G_1(\bm r_1,z_1;\bm r_3,\zeta)
G_0(\bm r_2,z_2;\bm r_4,\zeta)
{\mathcal A}(\bm r_3-\bm r_4),
\label{respo12}
\end{eqnarray}
as it follows from Eq. (\ref{respo1}). Here $G_1$ is determined by Eq. (\ref{delt1}).

For the case of the initial plane wave the first order corrections (\ref{phi1}) are analyzed in Appendices \ref{sec:firstphi},\ref{sec:smallk}. We give here the general estimates extracted from the analysis. For $k_0 r^2\ll z$ we find
\begin{eqnarray}
\varPhi_1(\bm r,z;\bm 0,z)
\sim C_n^4 k_0^{3-\mu} z^{3+\mu}, \quad
\nonumber \\
\varPhi_{1\alpha\beta}(\bm r,z;\bm 0,z)
\sim C_n^4 k_0^{4-\mu} z^{2+\mu}.
\label{gec4}
\end{eqnarray}
The quantities remain finite as $r\to0$. As it is demonstrated in Appendix \ref{sec:smallk}, for $k_0 r^2\gg z$ we arrive at the estimate
\begin{equation}
\varPhi_1\sim (C_n^2)^2 k_0^{5/2-\mu/2} r^{\mu-1}z^{7/2+\mu/2}.
\label{lug14}
\end{equation}
see Eq. (\ref{gec5}). Therefore $\varPhi_{1\alpha\beta}\propto r^{\mu-3}$.

Now we pass to the pair correlation function $\varXi$ (\ref{pairxi}). The first correction $\varXi_1$ to the expression (\ref{respo2}) consists of three terms
\begin{eqnarray}
\varXi_1=\varXi_1^{(1,3)}
+\varXi_1^{(2,2)}+\varXi_1^{(3,1)},
\label{phi5}
\end{eqnarray}
analogous to $\varPhi_1$. The term $\varXi_1^{(2,2)}$ is determined by the diagram depicted in Fig. \ref{fig:fdiagram4}. Two additional terms, $\varXi_1^{(1,3)}$ and $\varXi_1^{(3,1)}$, are related to the correction to the Green's function determined by the diagram depicted in Fig. \ref{fig:fdiagram2}. They can be thought as the correction to the Green's functions presented in the diagram for $\varXi_0$ depicted in Fig. \ref{fig:fdiagram1}.

The explicit expression for the contribution $\varXi_1^{(2,2)}$ is
\begin{eqnarray}
\varXi_1^{(2,2)}(\bm r_1,z_1;\bm r_2,z_2)=
\int d^2 r_3 d^2 r_4 d\zeta_3 d\zeta_4
\nonumber \\
G(\bm r_1,z_1;\bm r_3,z_3)
G^\star(\bm r_2,z_2;\bm r_4,z_4)
\Theta(\bm r_3,z_3;\bm r_4,z_4). \quad
\label{xi122} \\
\Theta(\bm r_3,z_3;\bm r_4,z_4)=
\frac{1}{2k_0^2} \varXi_{0,\alpha\beta}^2(\bm r_3,z_3;\bm r_4,z_4). \quad
\label{theta122}
\end{eqnarray}
The coefficients here are determined by the expressions (\ref{interaction},\ref{action2}). The contributions
$\varXi_1^{(1,3)}$, $\varXi_1^{(3,1)}$ are written as
\begin{eqnarray}
\varXi_1^{(1,3)}(\bm r_1,z_1; \bm r_2,z_2)
=k_0^2\int d^2 r_3\, d^2 r_4\, d\zeta\, C_n^2(\zeta)
\nonumber \\
G_0(\bm r_1,z_1;\bm r_3,\zeta)
G_1^\star(\bm r_2,z_2;\bm r_4,\zeta)
{\mathcal A}(\bm r_3-\bm r_4),
\label{respo213}
\end{eqnarray}
and
\begin{eqnarray}
\varXi_1^{(3,1)}(\bm r_1,z_1; \bm r_2,z_2)
=k_0^2\int d^2 r_3\, d^2 r_4\, d\zeta\, C_n^2(\zeta)
\nonumber \\
G_1(\bm r_1,z_1;\bm r_1,\zeta)
G_0^\star(\bm r_2,z_2;\bm r_4,\zeta)
{\mathcal A}(\bm r_3-\bm r_4),
\label{respo231}
\end{eqnarray}
as it follows from Eq. (\ref{respo2}). Here $G_1$ is, again, the first correction to the Green's function determined by Eq. (\ref{delt1}).

For the initial plane wave the corrections (\ref{phi5}) are analyzed in Appendices \ref{sec:firstphi},\ref{sec:smallk},\ref{sec:largek}. Note a remarkable cancellation in $\varXi$, occurring at small separations, which is analyzed in Appendix \ref{sec:largek}. A provenance of the cancellation will be specially discussed below. Based on these results, we give here only general estimates.

For $k_0 r^2\ll z$ we arrive at the estimates
\begin{eqnarray}
\varXi_1(\bm r,z;\bm 0,z)
\sim C_n^4 k_0^{3-\mu} z^{3+\mu}, \quad
\label{ppi1} \\
\varXi_{1\alpha\beta}(\bm r,z;\bm 0,z)
\sim C_n^4 k_0^{7/2-\mu/2} z^{5/2+\mu/2} r^{\mu-1}.
\label{phi15}
\end{eqnarray}
Contrary to Eq. (\ref{gec4}) the expression (\ref{phi15}) tends to infinity as $r\to 0$. The case $k_0 r^2\gg z$ is analyzed in Appendix \ref{sec:smallk}. As a result of the analysis, we arrive at the estimate
\begin{equation}
\varXi_1\sim (C_n^2)^2 k_0^{5/2-\mu/2} r^{\mu-1}z^{7/2+\mu/2}.
\label{lug16}
\end{equation}
see Eq. (\ref{gec11}).

Comparing the zeroth and the first approximations, one establishes the applicability condition of perturbation theory. Let us first consider small $r$, $k_0 r^2\ll z$. Then the behavior of the derivatives $\varPhi_{\alpha\beta}$ and $\varXi_{\alpha\beta}$ is quite different. In the zeroth approximation it is determined by Eqs. (\ref{vix3},\ref{vix4}). The first corrections $\varPhi_{1\alpha\beta}$, $\varXi_{1\alpha\beta}$ are estimated in accordance with Eqs. (\ref{gec4},\ref{phi15}). Then
\begin{eqnarray}
{\nabla^2\varPhi_1}/{\nabla^2\varPhi_{0}}
\sim {\nabla^2\varXi_{1}}/{\nabla^2\varXi_{0}}
\sim \sigma_R^2,
\label{res6}
\end{eqnarray}
where the Rytov dispersion is determined by Eq. (\ref{res67}).

Let us pass to large $r$, $k_0 r^2\gg z$. Then $\varPhi_{0\alpha\beta}$, $\varXi_{0\alpha\beta}$ are determined by Eqs. (\ref{vix3},\ref{larger}). The estimations for the dominant corrections to $\varPhi,\varXi$ are given by Eqs. (\ref{lug14},\ref{lug16}). Therefore we arrive at
\begin{eqnarray}
{\varPhi_1}/{\varPhi_{0}}
\sim {\varXi_{1}}/{\varXi_{0}}
\sim \sigma_R^2 \xi^{-1},
\label{res5} \\
\xi=k_0 r^2 /(4z).
\label{conc1}
\end{eqnarray}
Besides the dimensionless factor (\ref{res67}), figuring in Eq. (\ref{res6}), the ratio (\ref{res5}) contains the small factor $\xi^{-1}$.

We conclude that a smallness of the dimensionless parameter (\ref{res67}) ensures smallness of the corrections if $k_0 r^2\lesssim z$. For $k_0 r^2\gg z$ there appears the additional small factor $\xi^{-1}$ in the applicability condition of the perturbation series. Thus the smallness of the Rytov dispersion guarantees applicability of perturbation theory.

\subsection{Triple vertex}

Here we discuss corrections to the interaction vertex determined by the action (\ref{interaction}). By definition, the vertex function $\varGamma_{\alpha\beta}$ is introduced via the following correlation function
\begin{eqnarray}
\langle \eta(\bm r_1,z_1) \bar p(\bm r_2,z_2) \bar p(\bm r_3,z_3) \rangle =
\nonumber \\
\frac{1}{k_0}\int d^2r_4\, dz_4\, d^2r_5\, dz_5\, d^2r_6\, dz_6\,
\nonumber \\
\langle \eta(\bm r_1,z_1)\bar p(\bm r_4,z_4)\rangle
\varGamma_{\alpha\beta}(\bm r_5,z_5;\bm r_6,z_6;\bm r_4,z_4)
\nonumber \\
\langle \partial_\alpha \eta(\bm r_5,z_5)\bar p(\bm r_2,z_2)\rangle
\langle \partial_\beta \eta(\bm r_6,z_6)\bar p(\bm r_3,z_3)\rangle.
\label{trpl1}
\end{eqnarray}
The zero contribution to the vertex function $\varGamma_{\alpha\beta}$ is
\begin{eqnarray}
\varGamma_{0,\alpha\beta}(\bm r_1,z_1;\bm r_2,z_2;\bm r_3,z_3) =
\nonumber \\
\delta_{\alpha\beta}
\delta(\bm r_1-\bm r_3) \delta(\bm r_2-\bm r_3)
\delta(z_1-z_3)\delta(z_2-z_3),
\label{trpl2}
\end{eqnarray}
as it follows from Eq. (\ref{interaction}).

\begin{figure}
\begin{center}
\includegraphics[width=0.65 \textwidth]{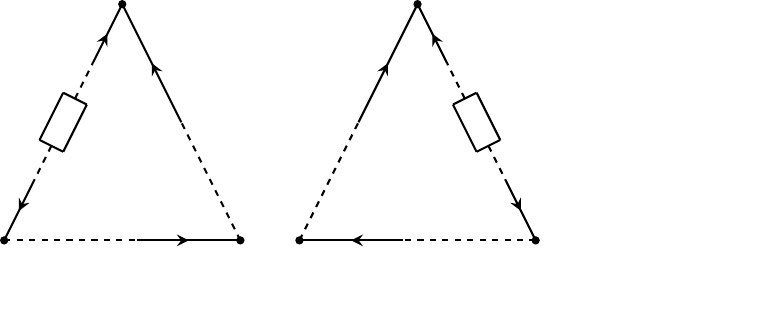}
\end{center}
\caption{Two additional first-order (one-loop) corrections to the triple vertex function.}
 \label{fig:fdiagram6}
 \end{figure}

The first corrections to the vertex function $\varGamma_{\alpha\beta}$ are determined by the Feynman diagrams depicted in Figs. \ref{fig:fdiagram5} and \ref{fig:fdiagram6}. Analytically, the corrections are written as
\begin{eqnarray}
\varGamma_{1,\alpha\beta}(\bm r_1,z_1;\bm r_2,z_2;\bm r_3,z_3) =
\nonumber \\
-\frac{1}{k_0^2}
\partial_\gamma G_0(3;1)
\partial_\gamma  G_0(3;2)
\varPhi_{0,\alpha\beta}(1;2)
\nonumber \\
-\frac{1}{2k_0^2}
\varPhi_{0,\alpha\gamma}(1;3)
\partial_\gamma  G_0(3;2)
\partial_\beta G_0(2;1)
\nonumber \\
-\frac{1}{2k_0^2}
\varPhi_{0,\beta\gamma}(2;3)
\partial_\gamma G_0(3;1)
\partial_\alpha G_0(1;2).
\label{trpl3}
\end{eqnarray}
Here we use the designations like $G(3;1)=G(\bm r_3,z_3;\bm r_1,z_1)$. One should substitute into Eq. (\ref{trpl3}) the expressions (\ref{respo1},\ref{logpl}).

We see that the correction (\ref{trpl3}) is determined only by the functions $G_0,\varPhi_{0,\alpha\beta}$. The assertion is valid for any order of the perturbation series. It is a consequence of existing a closed stochastic equation for $\eta$, see Eq. (\ref{log1}). As we established, $\varPhi_{0,\alpha\beta}$ has no singularities at small separations. Therefore $\varGamma_{1,\alpha\beta}$ has no singularities at small separations as well.

Taking all lateral separations in the expression (\ref{trpl3}) to be of the order of $\sqrt{z/k_0}$, we arrive at the estimate
\begin{eqnarray}
\varGamma_{1,\alpha\beta}
\sim C_n^2 k_0^{7/2-\mu/2} z^{\mu/2-5/2},
\label{trpl4}
\end{eqnarray}
see Eq. (\ref{vix4}). The estimate (\ref{trpl4}) means that $\varGamma_{1,\alpha\beta}$ has the same smallness (\ref{res67}) compared to $\varGamma_{0,\alpha\beta}$. Thus, we again face the assertion that the condition for the validity of the perturbation series is $\sigma_R^2 \ll 1$.


\section{Role of large-scale fluctuations of $\nu$}
\label{sec:largesc}

Let us perform the transformation $\ln\Psi\to u$ defined by the relations
\begin{eqnarray}
\ln\Psi=u(\bm r- \nabla \chi) +i k_0 \partial_z \chi,
\label{ori1} \\
\nu(\bm r,z)\to \nu(\bm r- \nabla \chi,z),
\label{ori2}
\end{eqnarray}
where $\chi$ is some function of $\bm r,z$. One can easily check that the equation (\ref{logpsi}) is invariant under the transformation (\ref{ori1},\ref{ori2}) if $\chi=\bm \alpha z \bm r$ where $\bm\alpha$ is a small vector. The property reflects the isotropy of the original wave equation, since for $\chi=\bm \alpha z \bm r$ the transformation (\ref{ori1},\ref{ori2}) is the rotation of the reference system by a small angle $\bm\alpha$.

Further we assume that the function $\chi$ is a slow function of the lateral coordinates $x,y$ and an arbitrary function of $z$. Then the equation for $u$ is written as
\begin{eqnarray}
\partial_z u
-\frac{i}{2k_0}[ \nabla^2 u +(\nabla u)^2]
-i k_0 \nu +ik_0 \partial_z^2 \chi
=\dots, \quad
\label{ori3}
\end{eqnarray}
where dots designate terms containing second or more space derivatives of $\chi$. The terms are small and can be neglected.

With the same accuracy, the correlation function (\ref{delta}) is kept at the transformation (\ref{ori2}) since the correction $\nabla\chi(\bm r_1)-\nabla \chi(\bm r_2)$ to the argument $\bm r=\bm r_1-\bm r_2$ of the function ${\mathcal A}$ is small compared to $\bm r$. Thus, in the main approximation, we arrive at the equation for the field $u$, which has the same structure as Eq. (\ref{logpsi}), but with an effective refractive index $\nu-\partial_z^2\chi$.

As we demonstrated calculating the correction $\varXi_1$ for small separations, the leading terms of the contributions (\ref{xi122},\ref{respo213},\ref{respo231}) cancel each other out, see Appendix \ref{sec:largek} for details. Now we can explain the provenance of the cancellation. The expressions (\ref{vxi181},\ref{gec141}) show that at large wave vectors $k$ (corresponding to small separations) the main terms (to be cancelled) are gained at small wave vectors $q$. By other words, the main terms are related to large-scale contributions to $\varPhi_0,\varXi_0$, which can be recognized inside the loops in Figs. \ref{fig:fdiagram2},\ref{fig:fdiagram4}. Therefore we should focus on the large-scale contribution to $\eta$, or, more precisely, on the large-scale contribution to $\ln \Psi$.

Let us choose the field $\chi$ in accordance with the definition
\begin{eqnarray}
\partial_z^2 \chi=
\int_{k<\kappa}\frac{d^2 k}{(2\pi)^2}
\tilde\nu(\bm k,z) [\exp(i\bm k \bm r)-1].
\label{ori4}
\end{eqnarray}
The relation means that we exclude the large-scale component in the effective refractive index $\nu-\partial_z^2\chi$.. The border wave vector $\kappa$ has to be chosen to satisfy the inequality $\kappa^2 \ll k_0/z$. The condition guarantees the smallness of the lateral derivatives of the field $\chi$.

The equation (\ref{ori4}) is the second-order ordinary differential equation. We take the solution with zero final conditions for $\chi$ and $\partial_z \chi$. Then at the final point $\ln\Psi=u$ as a consequence of Eq. (\ref{ori1}). The definition (\ref{ori4}) for the zero final conditions leads to the conclusion $\chi(\bm 0,z)=0$. Thus $u(\bm r,z)$ remains close to $\ln\Psi$ at least for distances $r\lesssim \kappa^{-1}$.

We proved that the field $u$ is close to $\ln\Psi$. Besides, there are no divergencies at small wave vectors in the perturbation series for the field $u$ determined by Eq. (\ref{ori3}) since the effective refractive index $\nu-\partial_z^2\chi$ have no wave vectors smaller than $\kappa$. Therefore, no divergences at small wave vectors should occur in the perturbation series for $\ln \Psi$. This property is demonstrated explicitly for the initial plane wave in Appendix \ref{sec:largek}.

The cancellations described above take place for correlation functions of the field $\eta$ with the same values of the argument $z$. However, these cancellations are not available for correlation functions with different propagation lengths. The most important such average is the Green's function, defined by Eq. (\ref{agra1}). Explicit calculations performed in Appendix \ref{sec:greenf1} demonstrate an increase in the first correction $\sim k^{1+\mu}$ to $\tilde{G}({\bm k}, z,0)$ with an increase in the wave vector, see Eq. (\ref{gone1}). Such growing contributions arise in the higher orders of perturbation theory, and in order to determine the asymptotic behavior of the Green's function $\tilde{G}({\bm k}, z,0)$ they must be summed up.

It is noted in Appendix \ref{sec:greenf1} that these increasing contributions are due to the interaction between small-scale and large-scale fluctuations with the wave vectors $q\sim k^{-1}(k_0/z)$. The effect of this interaction on the Green's function is taken into account in all orders of perturbation theory through the shift transformation (\ref{ori1},\ref{ori2}), with respect to which the Green's function is not invariant. Therefore for large wave vectors $k^2\gg k_0/z$ the Green's function is written as
\begin{equation}
\tilde{G}({\bm k}, z,0)\approx
\left\langle
\exp\left(-i\frac{{\bm k}^2}{2k_0}z-i{\bm k}{\bm \nabla}\chi(z)\right)\right\rangle.
\label{fga1}
\end{equation}
The border wave vector $\kappa$ is determined by smallness of the neglected terms: $\kappa \ll k^{-1}(k_0/z)$.

In the considered asymptotic case, the dependence on the vector ${\bm r}$ in the gradient of $\chi$ is irrelevant. Therefore, averaging over the random function ${\nabla \chi (z)}$, which has Gaussian statistics, can be performed explicitly. As a result, one finds
\begin{eqnarray}
\ln\tilde{G}({\bm k}, z,0)+i\frac{z{k}^2}{2k_0} \sim
-\left(\frac{z k^2}{k_0}\right)^{1/2+\mu/2}\sigma_R^2.
\label{fga2}
\end{eqnarray}
Despite the smallness of the parameter $\sigma_R^2$, the correction can become significant if $k^2 \gg k_0 / z$.

The expression (\ref{fga2}) indicates that the response to short-wavelength perturbations of the refractive index, as given by Eq. (\ref{agra1}), is suppressed due to turbulent fluctuations. Note that a similar effect caused by the interaction of large-scale eddies with small-scale ones occurs in hydrodynamic turbulence, see Ref. \cite{BL87}.

\section{Conclusion}
\label{sec:conclusion}

We considered peculiarities of a light wave propagating in a turbulent medium. Our focus is on analyzing the pair correlation function of phase derivatives, which is directly observable. For the purpose we developed the perturbation series for the logarithm of the wave packet envelope, $\ln \Psi$.

We have analytically derived the first corrections to the pair correlation functions of $\ln \Psi$, $\ln |psi^\star$ and for the vertex function. Based on the analysis of the corrections, we can assert that the condition under which the perturbation series is applicable is smallest of the Rytov dispersion $\sigma_R^2$, in accodance with the common believe.

Besides $\sigma_R^2$ there is an additional dimensionless parameter in the problem which is the ratio of the distance $r$ between the observation points and the radius of the first Fresnel zone $\sqrt{2\pi z/k_0}$. Remarkably, at the condition $r\gg \sqrt{2\pi z/k_0}$ the terms of the perturbation series contain some additional small factors which are powers of the ratio.

While developing the perturbation series, we encountered the cancellation of some leading terms, which could potentially limit the applicability of perturbation theory. These cancellations can be explained by the symmetry properties of the equation for $\ln \Psi$, which lie outside the scope of perturbation theory. This symmetry leads to certain conclusions about the role of large-scale fluctuations of the refractive index.

It should be stressed that a perturbation theory can be developed directly for the envelope $\Psi$. However, the corrections to the correlation functions of the fields $\Psi$ and $\Psi^*$ appear to grow with the distance between the points of observation, in contrast to the corrections to the correlation functions of $\ln\Psi$. As a result, even with a small value of the Rytov variance, $\sigma^2_R$, perturbation theory for the correlation functions of $\Psi$ and $\Psi^*$ becomes inapplicable at large enough distances $r$. The essential difference between the perturbation series for $\Psi$ and $\eta$ was previously noted in Ref. \cite{SV}, where a homogeneous perturbation of $\nu$ was considered.

It is well known that, under certain conditions, atmospheric turbulence may not follow the Kolmogorov model as discussed in Ref. \cite{Korot21} This is due to the breaking of the isotropy of the turbulence, for example. However, this phenomenon requires a more in-depth investigation, which is beyond the scope of this paper.

\acknowledgements

This work was supported by the scientific program of the National Center for Physics and Mathematics of RF, Section 4, stage 2026–2028. We thank F.A.Starikov for valuable discussions.



\appendix

\section{Zeroth approximation for the perturbed plane wave}
\label{sec:zeroth}

General expressions for the zeroth values of $\varPhi_0$, $\varXi_0$ are given by Eqs. (\ref{respo1},\ref{respo2}). Since for the initial plane wave the Green's function $G_0$ depends only on the difference of the lateral coordinates, then the functions (\ref{respo1},\ref{respo2}) are functions only of the difference $\bm r_1-\bm r_2$ as well. In this case it is convenient to write $\varPhi_0$, $\varPhi_0$ as the Fourier integrals
\begin{eqnarray}
\varPhi_0(\bm r_1,z_1;\bm r_2,z_2)=
\nonumber \\
\int \frac{d^2q}{(2\pi)^2} \tilde \varPhi_0(\bm q,z_1,z_2)
\exp(i\bm q \bm r),
\label{lig1} \\
\varXi_0(\bm r_1,z_1;\bm r_2,z_2)=
\nonumber \\
\int \frac{d^2 q}{(2\pi)^2} \exp(i \bm q \bm r)
\tilde\varXi_0(\bm q,z_1,z_2), \quad
\label{vxi1}
\end{eqnarray}
where $\bm r=\bm r_1-\bm r_2$.

The expressions for the Fourier transforms $\tilde\varPhi_0$, $\tilde\varXi_0$ can be found from Eqs. (\ref{respo1},\ref{respo2},\ref{logpl},\ref{log4}). They are
\begin{eqnarray}
\tilde \varPhi_0(\bm q,z_1,z_2)=-k_0^2 \tilde{\mathcal A}(q)
\nonumber \\
\int d\zeta\, C_n^2
\exp\left(-i\frac{z_1+z_2-2\zeta}{2k_0}q^2\right),
\label{lag1}
\end{eqnarray}
and
\begin{eqnarray}
\tilde\varXi_0(\bm q,z_1,z_2)
=k_0^2 \tilde{\mathcal A}(\bm q)
\int d\zeta\, C_n^2
\nonumber \\
\exp\left(-i\frac{z_1-z_2}{2k_0}q^2\right),
\label{vxi2}
\end{eqnarray}
where the integration over $\zeta$ goes from $0$ to $\min(z_1,z_2)$, and $\tilde{\mathcal A}$ is determined by Eq. (\ref{afo2}).

The integrals (\ref{lig1},\ref{vxi1}) diverge at small $q$, thus producing the contribution proportional to ${\mathcal A}_0$, see Eq. (\ref{corrfa2}), which is determined by the outer scale of turbulence. However, these large constants, which are independent of the lateral coordinates, are not observable and do not appear in the perturbation series due to the structure of the interaction term (\ref{interaction}). That is why we pass to examining the derivatives (\ref{qalbe3}).

For the derivatives of $\varPhi$ one obtains
\begin{eqnarray}
\varPhi_{0\alpha\beta}(\bm r_1,z_1;\bm r_2,z_2)
\nonumber \\
=\int \frac{d^2 q}{(2\pi)^2}
\exp(i\bm q \bm r)
\tilde\varPhi_{0\alpha\beta}(\bm q,z_1,z_2),
\label{lyg2}
\end{eqnarray}
where
\begin{eqnarray}
\tilde\varPhi_{0\alpha\beta}(\bm q,z_1,z_2)
=-2\pi{k_0^2}\Gamma(2+\mu) \sin(\pi\mu/2)
\nonumber \\
\int d\zeta\, C_n^2
\frac{q_\alpha q_\beta}{q^{\mu+3}}
\exp\left(-i\frac{z_1+z_2-2\zeta}{2k_0}q^2\right),
\label{lag2}
\end{eqnarray}
as a consequence of Eq. (\ref{lag1}). Analogously for the derivatives of $\varXi$ one obtains
\begin{eqnarray}
\varXi_{0\alpha\beta}(\bm r_1,z_1;\bm r_2,z_2)
\nonumber \\
=\int \frac{d^2 q}{(2\pi)^2}
\exp(i\bm q \bm r)
\tilde\varXi_{0\alpha\beta}(\bm q,z_1,z_2),
\label{vxi3}
\end{eqnarray}
where
\begin{eqnarray}
\tilde\varXi_{0\alpha\beta}(\bm q,z_1,z_2)
=2\pi\Gamma(2+\mu)\sin(\pi \mu/2) k_0^2
\frac{q_\alpha q_\beta}{q^{\mu+3}}
\nonumber \\
\int d\zeta\, C_n^2
\exp\left(-i\frac{z_1-z_2}{2k_0}q^2\right),
\label{vxi5}
\end{eqnarray}
as a consequence of the expression (\ref{afo2}).

The quantities (\ref{lyg2},\ref{vxi3}) remain finite as $\bm r_1-\bm r_2\to 0$. However, the estimates of the quantities are different. They are
\begin{eqnarray}
\varPhi_{0\alpha\beta}(\bm r_1,z_1;\bm r_1,z_2)\sim
C_n^2  k_0^{5/2-\mu/2} z^{1/2+\mu/2}, \quad
\label{vxi6} \\
\varXi(\bm r_1,z_1;\bm r_1,z_2)\sim
C_n^2 k_0^{5/2-\mu/2} z |z_1-z_2|^{\mu/2-1/2}. \quad
\label{vxi7}
\end{eqnarray}
The expressions (\ref{vxi6},\ref{vxi7}) follow from the estimates for the characteristic wave vectors $q$ determined by the oscillating exponents in Eqs. (\ref{lag2},\ref{vxi5}). The quantity (\ref{vxi7}) diverges as $z_1-z_2\to0$. The opposite limit where $z_1=z_2$ and the difference $\bm r_1-\bm r_2$ is small is also singular. In this case the characteristic wave vector $q\sim |\bm r_1-\bm r_2|^{-1}$ is determined by the exponent $\exp(i\bm q \bm r)$ in Eq. (\ref{vxi3}). The estimate for the characteristic wave vector leads to Eq. (\ref{vix3}).

\section{Correction $G_1$ to the Green's Function}
\label{sec:greenf1}

Here we specify the correction to the Green's Function for the case of the initial plane wave. Then the general relation (\ref{delt1}) can be rewritten in Fourier representation as
\begin{eqnarray}
\tilde G_1(\bm k,z_1,z_2)
=\int d\zeta_1 d\zeta_2
\tilde G_0(\bm k,z_1,\zeta_1)
\nonumber \\
\tilde\varSigma(\bm k,\zeta_1,\zeta_2)
\tilde G_0(\bm k,\zeta_2,z_2),
\label{delt2}
\end{eqnarray}
where
\begin{eqnarray}
\tilde\varSigma(\bm k,\zeta_1,\zeta_2)=
\frac1{k_0^2}\int \frac{d^2 q}{(2\pi)^2}
\tilde\varPhi_{0\alpha\beta}(\bm q,\zeta_1,\zeta_2)
\nonumber \\
\tilde G_0(\bm k-\bm q,\zeta_1,\zeta_2) (k_\alpha-q_\alpha) k_\beta.
\label{delt3}
\end{eqnarray}
The expression for $\tilde G_0$ is given by Eq. (\ref{log4}) and the expression for $\tilde\varPhi_{0\alpha\beta}$ is given by Eq. (\ref{lag2}).

We begin our analysis with the case $k^2\ll k_0/z$. Then the wave vector $q$ in Eq. (\ref{delt3}) is estimated as $\sqrt{k_0/z}$. Thus for small $k$ the inequality $k\ll q$ is satisfied. In the limit one finds from Eq. (\ref{delt3})
\begin{eqnarray}
\tilde\varSigma(\bm k,\zeta_1,\zeta_2)=
\frac{1}{k_0^2}
k_\alpha k_\beta
\int \frac{d^2 q}{(2\pi)^2}
\tilde\varPhi_{\alpha\beta}(\bm q)
\tilde G(\bm q)
\nonumber \\
+\frac{1}{k_0^2}
k_\beta k_\gamma
\int \frac{d^2 q}{(2\pi)^2}
\tilde\varPhi_{\alpha\beta}(\bm q)
\frac{\partial}{\partial q_\gamma}\tilde G(\bm q) q_\alpha.
\label{dolt1}
\end{eqnarray}
Taking here the integral over $q$, one obtains
\begin{eqnarray}
\tilde\varSigma(\bm k,\zeta_1,\zeta_2)=
\nonumber \\
-\frac{k^2}{4}\Gamma(2+\mu)\sin(\pi \mu/2)
\Gamma\left(\frac{1}{2}-\frac{\mu}{2}\right)
\left(\frac{i}{k_0}\right)^{\mu/2-1/2}
\nonumber \\
\int_0^{\zeta_2}d\zeta\, C_n^2
\left[1-(\zeta_1-\zeta_2)\frac{\partial}{\partial \zeta}\right]
(\zeta_1-\zeta)^{\mu/2-1/2}
\nonumber
\end{eqnarray}
If $C_n^2$ is independent of $z$ then
\begin{eqnarray}
\tilde\varSigma(\bm k,\zeta_1,\zeta_2)=
-\frac{1}{4}\Gamma(2+\mu)\sin(\pi \mu/2)
\nonumber \\
\Gamma(1/2-\mu/2) C_n^2 k^2
\left(\frac{i}{k_0}\right)^{\mu/2-1/2}
\nonumber \\
\left\{\frac{\mu+3}{\mu+1}
\left[\zeta_1^{\mu/2+1/2}
-(\zeta_1-\zeta_2)^{\mu/2+1/2}\right]
\right. \nonumber \\ \left. \phantom{\frac{2}{3}}
-\zeta_2\zeta_1^{\mu/2-1/2} \right\}.\qquad
\label{dolt2}
\end{eqnarray}
Now we proceed to calculating $G_1$ from Eq. (\ref{delt2}). For small $k$ one can substitute in Eq. (\ref{delt2}) both $\tilde G_0$ by unities to obtain after integration over $\zeta_1$ and $\zeta_2$
\begin{eqnarray}
\tilde G_1(\bm k,z_1,z_2)=
-\frac{1}{4}\Gamma(2+\mu)\sin(\pi \mu/2)
\nonumber \\
\Gamma(1/2-\mu/2) C_n^2 k^2
\left(\frac{i}{k_0}\right)^{\mu/2-1/2}
\nonumber \\
(z_1-z_2)^2\frac{(5+\mu)z_1^{1/2+\mu/2}-4(z_1-z_2)^{1/2+\mu/2}}{(1+\mu)(5+\mu)}.
\label{dolt22}
\end{eqnarray}
Note that the quantity tends to zero as $z_1-z_2\to 0$.

Let us pass to the case of large $k$, $k^2\gg k_0/z$. The integration over $\bm q$ in Eq. (\ref{delt3}) is restricted by $q\lesssim k_0/(kz)$. Therefore $q\ll k$ in the case. Neglecting $q$ in comparison with $k$ everywhere it is possible we obtain
\begin{eqnarray}
\tilde G_1(\bm k,z_1,z_2)=
-\exp\left(-i\frac{z_1-z_2}{2k_0}k^2\right)
\int d\zeta_1\, d\zeta_2
\nonumber \\
\int \frac{d^2 q}{(2\pi)^2} k_\alpha k_\beta
\int_0^{\zeta_2} d\zeta\, C_n^2
\exp\left(i\frac{\zeta_1-\zeta_2}{k_0}\bm q \bm k\right)
\nonumber \\
2\pi\Gamma(2+\mu) \sin(\pi\mu/2)
\frac{q_\alpha q_\beta}{q^{\mu+3}}, \quad
\label{gon21}
\end{eqnarray}
where we substituted Eqs. (\ref{log4},\ref{lag2}). The oscillating exponent in Eq. (\ref{gon21}) explains the restriction $q\lesssim k_0/(kz)$ for the actual integration region. The integration over $\zeta_1,\zeta_2$ in Eq. (\ref{gon21}) is performed in the intervals $z_2<\zeta_2<\zeta_1<z_1$. Therefore the quantity (\ref{gon21}) tends to zero as $z_1-z_2\to0$. Using the definitions (\ref{afo2},\ref{parpara}), one finds from Eq. (\ref{gon21})
\begin{eqnarray}
\tilde G_1(\bm k,z_1,z_2)=
-\exp\left(-i\frac{z_1-z_2}{2k_0}k^2\right)
\int d\zeta_1\, d\zeta_2
\nonumber \\
k_\alpha k_\beta S_{\alpha\beta}\left(\frac{\zeta_1-\zeta_2}{k_0} \bm k\right)
\int_0^{\zeta_2} d\zeta\, C_n^2. \quad
\label{gon22}
\end{eqnarray}
If $C_n^2=\mathrm{const}$ then the integrals in Eq. (\ref{gon22}) can be taken explicitly and one obtains
\begin{eqnarray}
\tilde G_1(\bm k,z_1,z_2)=
\exp\left(-i\frac{z_1-z_2}{2k_0}k^2\right) C_n^2
\nonumber \\
\frac{1}{(\mu+1)(\mu+2)}(z_1-z_2)^{\mu+1}[z_1+(\mu+1)z_2]
\nonumber \\
k^{1+\mu} k_0^{1-\mu} \sqrt\pi \frac{\Gamma(1/2-\mu/2)}{\Gamma(-\mu/2)}  , \quad
\label{gone1}
\end{eqnarray}
where we used Eq. (\ref{parpamu}).

\section{First corrections $\varPhi_1$, $\varXi_1$ for the initial plane wave}
\label{sec:firstphi}

Here we analyze the corrections $\varPhi_1$, $\varXi_1$ for the case of the initial plane wave. Then all correlation functions are homogeneous in the lateral plane. Therefore it is instructive to pass to Fourier transforms over the differences of the lateral coordinates in the general expressions (\ref{phi122},\ref{respo11},\ref{respo12}).

Fourier transform $\tilde\varPhi_1^{(2,2)}$ of the expression (\ref{phi122}) is
\begin{eqnarray}
\tilde\varPhi_1^{(2,2)}(\bm k,z_1,z_2)
=-\int_0^{z_1} d\zeta_1 \int_0^{z_2} d\zeta_2
\nonumber \\
\tilde G_0(\bm k,z_1,\zeta_1)
\tilde G_0(\bm k,z_2,\zeta_2)
\tilde\varPi(\bm k,\zeta_1,\zeta_2). \quad
\label{dlt2} \\
\tilde\varPi(\bm k,z_1,z_2)=
\frac{1}{2k_0^2}\int \frac{d^2 q}{(2\pi)^2}
\tilde\varPhi_{0\alpha\beta}(\bm q,z_1,z_2)
\nonumber \\
\tilde\varPhi_{0\alpha\beta}(\bm k-\bm q,z_1,z_2).
\label{log14}
\end{eqnarray}
Substituting into the formulas (\ref{dlt2},\ref{log14}) the expressions (\ref{log4},\ref{lag2}), one obtains a multiple integral with an explicitly written integrand.

Now we pass to $\varPhi_1^{(1,3)}+\varPhi_1^{(3,1)}$. In Fourier representation
\begin{eqnarray}
\tilde \varPhi_1^{(3,1)}(\bm k,z_1,z_2)+
\tilde \varPhi_1^{(1,3)}(\bm k,z_1,z_2)
\nonumber \\
=-k_0^2 \int d\zeta C_n^2
\left[\tilde G_1(\bm k,z_1,\zeta)
\tilde {\mathcal A} (\bm k)
\tilde G_0(\bm k,z_2,\zeta) \right.
\nonumber \\ \left.
+\tilde G_0(\bm k,z_1,\zeta)
\tilde {\mathcal A} (\bm k)
\tilde G_1(\bm k,z_2,\zeta) \right],
\label{gec1}
\end{eqnarray}
where the integration over $\zeta$ goes from $0$ to $\min(z_1,z_2)$. Substituting here the expressions (\ref{afo2},\ref{log4},\ref{lag2},\ref{delt2},\ref{delt3}), one finds a multiple integral with an explicitly written integrand.

We are interested in
\begin{eqnarray}
\varPhi_1(\bm r,z;\bm 0,z)
=\int \frac{d^2 k}{(2\pi)^2} \exp(i\bm k \bm r)
\nonumber \\
\left[\varPhi_1^{(1,3)}(\bm k,z,z)
+\varPhi_1^{(2,2)}(\bm k,z,z)
+\varPhi_1^{(3,1)}(\bm k,z,z) \right].
\label{gec3}
\end{eqnarray}
Though this is a complicated multiple integral, it allows for an analysis of two limit cases of small $r$, $r^2\ll z/k_0$, and of large $r$, $r^2\gg z/k_0$.

For small $r$ the integral (\ref{gec3}) is determined by $k\sim\sqrt{k_0/z}$ and we find the estimates
\begin{eqnarray}
\varPhi_1(\bm r,z;\bm 0,z)
\sim C_n^4 k_0^{3-\mu} z^{3+\mu}, \quad
\nonumber \\
\varPhi_{1\alpha\beta}(\bm r,z;\bm 0,z)
\sim C_n^4 k_0^{4-\mu} z^{2+\mu}.
\nonumber
\end{eqnarray}
The quantities remain finite as $r\to0$.

Next we analyze the corrections (\ref{phi5}) for the case of the initial plane wave. Then all correlation functions are homogeneous in the lateral plane. Therefore it is instructive to pass to Fourier transforms over the differences of the lateral coordinates in the expressions (\ref{xi122},\ref{respo213},\ref{respo231}).

Fourier transform of $\varXi_1^{(2,2)}$ (\ref{xi122}) is written as
\begin{eqnarray}
\tilde\varXi_1^{(2,2)}(\bm k,z_1,z_2)=
\frac{1}{2 k_0^2}
\int d\zeta_1 d\zeta_2
\nonumber \\
\tilde\varTheta(\bm k,\zeta_1,\zeta_2)
\tilde G_0(\bm k,z_1,\zeta_1)
\tilde G_0^\star(\bm k,z_2,\zeta_2),
\label{vxi17} \\
\tilde\varTheta(\bm k,z_1,z_2)=
\frac{1}{2k_0^2}\int \frac{d^2 q}{(2\pi)^2}
\tilde\varXi_{0\alpha\beta}(\bm q,z_1,z_2)
\nonumber \\
\tilde\varXi_{0\alpha\beta}(\bm k+\bm q,z_1,z_2),
\label{vxi14}
\end{eqnarray}
performed in terms of $\bm r=\bm r_1-\bm r_2$. The quantities entering Eqs. (\ref{vxi17},\ref{vxi14}) are given by Eqs. (\ref{log4},\ref{vxi5}).

Fourier transform of the sum $\varXi_1^{(1,3)}+\varXi_1^{(3,1)}$ determined by Eqs. (\ref{respo213},\ref{respo231}) is written as
\begin{eqnarray}
\tilde \varXi_1^{(3,1)}(\bm k,z_1,z_2)+
\tilde \varXi_1^{(1,3)}(\bm k,z_1,z_2)
\nonumber \\
=-k_0^2 \int d\zeta C_n^2
\left[\tilde G_1(\bm k,z_1,\zeta)
\tilde {\mathcal A} (\bm k)
\tilde G_0^\star(\bm k,z_2,\zeta) \right.
\nonumber \\ \left.
+\tilde G_0(\bm k,z_1,\zeta)
\tilde {\mathcal A} (\bm k)
\tilde G_1^\star(\bm k,z_2,\zeta) \right],
\label{phi9}
\end{eqnarray}
where the integration over $\zeta$ goes from $0$ to $\min(z_1,z_2)$.

The complete correction is
\begin{eqnarray}
\varXi_1(\bm r,z;\bm 0,z)
=\int \frac{d^2 k}{(2\pi)^2} \exp(i\bm k \bm r)
\nonumber \\
\left[\varXi_1^{(1,3)}(\bm k,z,z)
+\varXi_1^{(2,2)}(\bm k,z,z)
+\varXi_1^{(3,1)}(\bm k,z,z) \right].
\label{phi10}
\end{eqnarray}
Unlike $\varPhi_1$, the consideration of the behavior of $\varXi_1$ at small $r$ is complicated. Its analysis is presented in Appendix \ref{sec:largek}.

\section{Large separations}
\label{sec:smallk}

Here, we present explicit expressions for the first corrections to the correlation functions $\varPhi$ and $\varXi$ at large separations $r$ between the points, $r^2 \gg z/k_0$. The separations correspond to small wave vectors $k$, $k^2\ll k_0/z$, for the Fourier components $\tilde\varPhi$ and $\tilde\varXi$ of the correlation functions. We discuss gradients of the correlation functions as well.

Let us analyze $\tilde\varPi(k)$ for small $k$, $k^2\ll k_0/z$. Then the characteristic $q$ in the integral (\ref{log14}) is estimated as $q\sim k$. Therefore the exponent in the expression (\ref{lag2}) can be put to unity and we find
\begin{equation}
\tilde \varPhi_0=-k_0^2 C_n^2 z \tilde {\mathcal A},
\nonumber
\end{equation}
for $C_n^2=\mathrm{const}$. Substituting the expression and Eq. (\ref{afo2}) into Eq. (\ref{log14}) and performing the integration over $\bm q$ one finds
\begin{eqnarray}
\tilde\varPi(\bm k,z_1,z_2)=
\frac{\pi}{2}k_0^2 z^2 \frac{1+\mu^2}{(1+\mu)^2}C_n^4k^{-2\mu}
\nonumber \\
\left[\Gamma(2+\mu) \sin(\pi\mu/2)
\frac{ \Gamma(1/2-\mu/2)}{\Gamma(1/2+\mu/2)}\right]^2
\frac{ \Gamma(\mu)}{\Gamma(1-\mu)},
\label{log12}
\end{eqnarray}
where $z=\min(z_1,z_2)$. For small $k$ the Green's functions $\tilde G_0$ in Eq. (\ref{dlt2}) can be substituted by unities and one finds from Eq. (\ref{log12})
\begin{eqnarray}
\tilde\varPhi_1^{(2,2)}(\bm k,z_1,z_2)=
-\frac{\pi}{12}k_0^2 z^4 \frac{1+\mu^2}{(1+\mu)^2}C_n^4k^{-2\mu}
\nonumber \\
\left[\Gamma(2+\mu) \sin(\pi\mu/2)
\frac{ \Gamma(1/2-\mu/2)}{\Gamma(1/2+\mu/2)}\right]^2
\frac{ \Gamma(\mu)}{\Gamma(1-\mu)},
\label{lig12}
\end{eqnarray}
if $C_n^2=\mathrm{const}$. Performing Fourier transform of Eq. (\ref{lig12}), one finds the expression
\begin{eqnarray}
\varPhi_1^{(2,2)}(\bm r,z;\bm 0,z)=
-\frac{k_0^2 z^4}{12\cdot 2^{2\mu}} \frac{1+\mu^2}{(1+\mu)^2}C_n^4r^{2\mu-2}
\nonumber \\
\left[\Gamma(2+\mu) \sin(\pi\mu/2)
\frac{ \Gamma(1/2-\mu/2)}{\Gamma(1/2+\mu/2)}\right]^2,
\label{lug12}
\end{eqnarray}
valid if $r^2\gg z/k_0$.

Substituting the expressions (\ref{afo2},\ref{log4}) into Eq. (\ref{gec1}) one obtains
\begin{eqnarray}
\tilde\varPhi_1^{(1,3)}
+\tilde\varPhi_1^{(3,1)}=
-2\int_0^z d\zeta\,
\exp\left[ -\frac{i(z-\zeta)}{2k_0}k^2 \right]
\nonumber \\
\tilde G_1(\bm k,z,\zeta)
2\pi \Gamma(2+\mu)\sin(\pi \mu/2)
k^{-3-\mu}k_0^2 C_n^2,
\label{gec2}
\end{eqnarray}
where we took the coinciding values $z_1=z_2=z$. For large $r$, $r^2\gg z/k_0$, the integral (\ref{gec3}) is determined by $k\sim r^{-1}\ll \sqrt{k_0/z}$. Therefore one can use the expression (\ref{dolt22}) to obtain from Eq. (\ref{gec2})
\begin{eqnarray}
\tilde\varPhi_1^{(1,3)}
+\tilde\varPhi_1^{(3,1)}=
i^{\mu/2-1/2}
\nonumber \\
\left[\Gamma(2+\mu)\sin(\pi \mu/2)\right]^2
k^{-1-\mu}k_0^{5/2-\mu/2} C_n^4,
\nonumber \\
\Gamma(1/2-\mu/2)
\frac{(11+\mu)\pi}{3(5+\mu)(7+\mu)} z^{7/2+\mu/2},
\nonumber
\end{eqnarray}
where we took the integral over $\zeta$. Performing Fourier transform, one finds
\begin{eqnarray}
\varPhi_1^{(1,3)}(\bm r,z;\bm 0,z)
+\varPhi_1^{(3,1)}(\bm r,z;\bm 0,z)
\nonumber \\
= i^{\mu/2-1/2}
k_0^{5/2-\mu/2}z^{7/2+\mu/2}r^{\mu-1}
\nonumber \\
\left[\Gamma(2+\mu)\sin(\pi \mu/2)
 C_n^2\right]^2
\nonumber \\
\frac{11+\mu}{3\cdot 2^{\mu+1} (5+\mu)(7+\mu)}
\frac{\Gamma^2(1/2-\mu/2)}{\Gamma(1/2+\mu/2)},
\label{gec5}
\end{eqnarray}
for $r^2\gg z/k_0$. For $\mu<1$ the contribution (\ref{gec5}) is much larger than the contribution (\ref{lug12}). Therefore Eq. (\ref{gec5}) determines the main contribution to $\varPhi_1$. The quantity $\varPhi_{1\alpha\beta}(\bm r,z;\bm 0,z)$ is obtained from Eq. (\ref{gec5}) by the substitution
\begin{equation}
r^{\mu-1}\to (1-\mu)
\left[\delta_{\alpha\beta}
-(3-\mu)\frac{r_\alpha r_\beta}{r^2}\right]
r^{\mu-3},
\label{gecc}
\end{equation}
derived by taking the derivatives.

For small $k$, $k^2\ll k_0/z$, one can use the asymptotic expression (\ref{vix2}) to obtain from Eq. (\ref{theta122})
\begin{eqnarray}
\varTheta(\bm r_1,z_1;\bm r_2,z_2)
=\frac{k_0^2}{2^{1+2\mu}}\frac{1+\mu^2}{(1+\mu)^2}\frac{1}{r^{2-2\mu}}
\nonumber \\
\left[{\Gamma(2+\mu)\sin(\pi \mu/2)}\right]^2
\nonumber \\
\left[\int_0^{\min(z_1,z_2)} d\zeta\, C_n^2\right]^2
\left[\frac{ \Gamma(1/2-\mu/2)}{\Gamma(1/2+\mu/2)}\right]^2,
\nonumber
\end{eqnarray}
where $\bm r=\bm r_1-\bm r_2$. Therefore one finds after Fourier transform
\begin{eqnarray}
\tilde\varTheta(\bm k,z_1,z_2)
=\frac{\Gamma(\mu)}{\Gamma(1-\mu)}
\frac{\pi k_0^2}{2 k^{2\mu}}\frac{1+\mu^2}{(1+\mu)^2}
\nonumber \\
\left[{\Gamma(2+\mu)\sin(\pi \mu/2)}\right]^2
\nonumber \\
\left[\int_0^z d\zeta\, C_n^2\right]^2
\left[\frac{ \Gamma(1/2-\mu/2)}{\Gamma(1/2+\mu/2)}\right]^2.
\label{vxi13}
\end{eqnarray}
For small $k$, $k^2\ll k_0/z$, the factors $\tilde G,\tilde G^\star$ in Eq. (\ref{vxi17}) can be substituted by unities. If $C_n^2=\mathrm{const}$ then we find from Eq. (\ref{vxi13})
\begin{eqnarray}
\tilde\varXi_1^{(2,2)}(\bm k,z,z)
=\frac{\Gamma(\mu)}{\Gamma(1-\mu)}
\frac{\pi k_0^2 z^4}{12 k^{2\mu}}\frac{1+\mu^2}{(1+\mu)^2}C_n^4
\nonumber \\
\left[{\Gamma(2+\mu)\sin(\pi \mu/2)}\right]^2
\left[\frac{ \Gamma(1/2-\mu/2)}{\Gamma(1/2+\mu/2)}\right]^2,
\label{vxi19}
\end{eqnarray}
analogously to Eq. (\ref{lig12}). Performing Fourier transform, one finds the expression
\begin{eqnarray}
\varXi_1^{(2,2)}(\bm r,z;\bm 0,z)=
\frac{k_0^2 z^4}{12\cdot 2^{2\mu}} \frac{1+\mu^2}{(1+\mu)^2}C_n^4r^{2\mu-2}
\nonumber \\
\left[\Gamma(2+\mu) \sin(\pi\mu/2)
\frac{ \Gamma(1/2-\mu/2)}{\Gamma(1/2+\mu/2)}\right]^2,
\label{lug13}
\end{eqnarray}
valid if $r^2\gg z/k_0$.

For small $k$, $k^2\ll k_0/z$, one finds using Eq. (\ref{dolt22})
\begin{eqnarray}
\tilde\varXi_1^{(1,3)}(\bm k,z,z) + \tilde\varXi_1^{(3,1)}(\bm k,z,z)
\nonumber \\
= -\frac{k_0^{5/2-\mu/2}}{k^{1+\mu}} \mathrm{Re}\,
\left[\Gamma(2+\mu)\sin(\pi \mu/2) C_n^2\right]^2
\nonumber \\
\Gamma(1/2-\mu/2)  i^{\mu/2-1/2}
\frac{\pi (11+\mu) z^{7/2+\mu/2}}{3(5+\mu)(7+\mu)}. \quad
\label{gec10}
\end{eqnarray}
For large $r$, $r^2\gg k_0/z$, we find from Eqs. (\ref{phi10},\ref{gec10})
\begin{eqnarray}
\varXi_1^{(1,3)}(\bm r,z;\bm 0,z) + \varXi_1^{(3,1)}(\bm r,z;\bm 0,z)
\nonumber \\
=-\mathrm{Re}\, i^{\mu/2-1/2}
\left[\Gamma(2+\mu)\sin(\pi \mu/2) C_n^2\right]^2
\nonumber \\
\frac{k_0^{5/2-\mu/2}}{2^{1+\mu}r^{1-\mu}}
\frac{\Gamma^2(1/2-\mu/2)}{\Gamma(1/2+\mu/2)}
\frac{(11+\mu) z^{7/2+\mu/2}}{3(5+\mu)(7+\mu)},
\label{gec11}
\end{eqnarray}
analogously to Eq. (\ref{gec5}). The quantity $\varXi_{1\alpha\beta}(\bm r,z;\bm 0,z)$ is obtained from Eq. (\ref{gec11}) by the substitution (\ref{gecc}). For $\mu<1$ the contribution (\ref{gec11}) is much larger than the contribution (\ref{lug13}). Therefore Eq. (\ref{gec11}) determines the main contribution to $\varXi_1$

\section{Small separations in $\varXi_1$.}
\label{sec:largek}

Here we analyze the first correction to the correlation functions $\varXi$, see Eq. (\ref{phi10}), and establish its behavior at small separations $r$, $r^2\ll z/k_0$. In terms of its Fourier transform, the case corresponds to large $k$, $k^2\gg k_0/z$.

Substituting the expressions (\ref{log4},\ref{vxi5}) into Eq. (\ref{vxi14}), one obtains from Eq. (\ref{vxi17})
\begin{eqnarray}
\tilde\varXi_1^{(2,2)}(\bm k,z,z)=k_0^2
\mathrm{Re}
\nonumber \\
\int_0^{z} d\zeta_1 \int_0^{\zeta_1}d\zeta_2\, \zeta_2^2
\int {d^2 q}
\frac{(\bm q \bm p)^2}{q^{\mu+3}p^{\mu+3}}
\nonumber \\
\left[\Gamma(2+\mu)\sin(\pi \mu/2) C_n^2\right]^2
\nonumber \\
\exp\left[-i\frac{\zeta_1-\zeta_2}{2k_0}(q^2+p^2-k^2)\right]
\label{vxi88}
\end{eqnarray}
where $\bm k=\bm p+\bm q$, and we assumed $C_n^2=\mathrm{const}$. Changing the order of integration over $\zeta_1, \zeta_2 $ and performing then the integration over $\zeta_1$ we arrive at the expression:
\begin{eqnarray}
\tilde\varXi_1^{(2,2)}(\bm k,z,z)=2 k_0^4
\mathrm{Re}
\nonumber \\
\int_0^{z} d\zeta \, (z-\zeta)
\int
\frac{d^2 {\bm q}}{q^{\mu+3}p^{\mu+3}}
\nonumber \\
\left[\Gamma(2+\mu)\sin(\pi \mu/2) C_n^2\right]^2
\nonumber \\
\left(1-
\exp\left[-\frac{i}{k_0}{\bm q}({\bm q - \bm k})\zeta\right]\right).
\label{vxi181}
\end{eqnarray}

In accordance with Eqs. (\ref{afo2},\ref{phi9})
\begin{eqnarray}
\tilde\varXi_1^{(1,3)}(\bm k,z,z)+\tilde\varXi_1^{(3,1)}(\bm k,z,z)
\nonumber \\
= 2k_0^2\,\mathrm{Re}\,
\int_0^z d\zeta\,
\exp\left[i\frac{z-\zeta}{2k_0}k^2\right]
\nonumber \\
\tilde G_1(\bm k,z,\zeta)
2\pi \Gamma(2+\mu)\sin(\pi \mu/2) C_n^2
k^{-3-\mu}
\label{gec7}
\end{eqnarray}
Substituting the expressions (\ref{log4},\ref{delt2},\ref{delt3}) into Eq. (\ref{gec7}) and performing integration over $\zeta_2$, one finds
\begin{eqnarray}
&&\tilde\varXi_1^{(1,3)}(\bm k,z,z)+\tilde\varXi_1^{(3,1)}(\bm k,z,z) =
\nonumber \\
&& -2k_0^4 \mathrm{Re}
\int \frac{d^2 q}{k^{3+\mu}q^{\mu+3}}
\left\{ z \int_0^z d\zeta \, \exp\left(-\frac{i\zeta}{k_0}q^2\right)-\right.
\nonumber \\
&&\left. \int_0^z d\zeta \left[z-\zeta+\int\limits_0^{z-\zeta} d\zeta_1 \,  \exp\left(-\frac{i}{k_0}q^2\zeta_1\right)\right]
\right.
\nonumber \\
&& \left.\!\exp\left[-\frac{i\zeta}{k_0}{\bm q}({\bm q - \bm k})\right]\right\}
\left[\Gamma(2+\mu) \sin(\pi\mu/2) C_n^2\right]^2 \! .
\label{gec141}
\end{eqnarray}
The correction $\tilde\varXi_1$ is the sum of the terms (\ref{vxi181}) and (\ref{gec141}).

For $k^2\gg k_0/z$ the leading contribution to the integral in Eq. (\ref{vxi181}) is gained from the domains of where $q\sim 1/k$ and $|\bm q - \bm k|\sim 1/k$. This domains give equal contributions which are proportional to $k^{-2}$. However, the sum of the contributions is cancelled by the leading contribution to the integral (\ref{gec141}). One can easily check the cancellation considering the domain of small $q$ in Eqs. (\ref{vxi181}) and (\ref{gec141}), neglecting $q$ comparing to $k$ and doubling the contribution from Eq. (\ref{vxi181}) because of the second relevant domain of integration where $\bm q - \bm k$ is small.

Thus, the main contribution to $\varXi_1$ is gained from the domains where $q^2\sim k_0/z$ or $(\bm q - \bm k)^2\sim k_0/z$. In this case the exponential $\exp(i\zeta {\bm k \bm q}/k_0)$ in Eqs. (\ref{vxi181}) and (\ref{gec141}) oscillates fast, and the corresponding terms can be neglected. Therefore we arrive at the following main contribution
\begin{eqnarray}
&&\tilde\varXi_1(\bm k,z,z)\approx
 2z k_0^4 \left[\Gamma(2+\mu) \sin(\pi\mu/2) C_n^2\right]^2
\nonumber\\
&&\mathrm{Re} \int \frac{d^2 q}{k^{3+\mu}q^{\mu+3}}
\left[ z -  \int_0^z d\zeta \, \exp\left(-\frac{i\zeta}{k_0}q^2\right)\right].
\nonumber
\end{eqnarray}
Taking here the integrals over $\bm q$ and $\zeta$, one finds the expression
\begin{eqnarray}
\tilde\varXi_1(\bm k,z,z)\approx
\cos\left[\frac{\pi (\mu+1)}{4}\right]
\nonumber \\
\frac{2 k_0^{7/2-\mu/2} z^{5/2+\mu/2} \Gamma\left({1}/{2}-{\mu}/{2}\right)}{k^{3+\mu}\pi (\mu+1)(\mu+3)},
\label{lk10}
\end{eqnarray}
correct at $k^2\gg k_0/z$.

After substituting Eq. (\ref{lk10}) into Eq. (\ref{phi10}) we arrive at the integral diverging at small $k$ where Eq. (\ref{lk10}) is unapplicable. However, being interested in the $r$-dependent part of $\varXi_1$, one can substitute $\exp(i \bm k \bm r)\to \exp(i \bm k \bm r)-1$ in Eq. (\ref{phi10}), thus subtracting from the expression an $r$-independent constant. After the subtraction the integral over $\bm k$ in Eq. (\ref{phi10}) becomes converging. Taking this integtal, one finds the following $r$-dependent contribution to $\varXi_1$
\begin{eqnarray}
\varXi_1\to - r^{\mu+1} z^{5/2+\mu/2}
k_0^{7/2-\mu/2}
\nonumber \\
\frac{2^{-\mu}  \Gamma(1/2-\mu/2)}{(1+\mu)^2 \Gamma(1/2+\mu/2)}
\nonumber \\
\cos\left[\frac{1}{4}(1+\mu)\pi\right] \Gamma(-3/2-\mu/2)
\nonumber \\
\left[\Gamma(2+\mu) \sin(\pi\mu/2) C_n^2\right]^2,
\label{gec25}
\end{eqnarray}
determined by $k\sim r^{-1}$. This is the main contribution to the $r$-dependent part of $\varXi_1$ at small $r$, $r^2\ll z/k_0$.

Calculating the derivatives of Eq. (\ref{gec25}),
\begin{equation}
\varXi_{1\alpha\beta}
\equiv -\frac{\partial^2}{\partial r_\alpha \partial r_\beta}
\varXi_1,
\nonumber
\end{equation}
one finds the explicit expression
\begin{eqnarray}
\varXi_{1\alpha\beta}= z^{5/2+\mu/2}
k_0^{7/2-\mu/2}
\nonumber \\
\left[\delta_{\alpha\beta}+(\mu-1)\frac{r_\alpha r_\beta}{r^2}\right] r^{\mu-1}
\nonumber \\
\frac{2^{-\mu}  \Gamma(1/2-\mu/2)}{(1+\mu) \Gamma(1/2+\mu/2)}
\nonumber \\
\cos\left[\frac{1}{4}(1+\mu)\pi\right] \Gamma(-3/2-\mu/2)
\nonumber \\
\left[\Gamma(2+\mu) \sin(\pi\mu/2) C_n^2\right]^2.
\label{gec26}
\end{eqnarray}
The correction (\ref{gec26}), valid at small $r$, has the same structure as the zero approximation (\ref{vix2}), including its $r$-dependence.

\end{document}